\documentstyle[11pt]{article}
\newcommand{\apm}{\!\stackrel{\leftrightarrow\;}{\partial_{\mu}}\!}
\newcommand{\apn}{\!\stackrel{\leftrightarrow\;}{\partial'_{\nu}}\!}

\newcommand{\cint}{\int_{\Sigma}d\Sigma}
\newcommand{\gapm}{\!\stackrel{\leftrightarrow\;}{D_{\mu}}\!}
\newcommand{\gapms}{\!\stackrel{\leftrightarrow\;}{D^*_{\mu}}\!}
\newcommand{\gapn}{\!\stackrel{\leftrightarrow\;}{D'_{\nu}}\!}
\newcommand{\gapns}{\!\stackrel{\leftrightarrow\;}{D'^*_{\nu}}\!}
\title{The general-covariant and gauge-invariant 
theory of quantum particles in classical backgrounds}
\author{Hrvoje Nikoli\'c \\
Theoretical Physics Division, Rudjer Bo\v{s}kovi\'{c} Institute, \\
P.O.B. 180, HR-10002 Zagreb, Croatia \\
{\normalsize hrvoje@faust.irb.hr} \\
\makebox[1in]{} \\
}
\date{\today}
\begin{document}
\maketitle
\begin{abstract}
A new approach to the concept of particles and their production in quantum
field theory is developed.
A local operator describing the current of particle density is 
constructed for scalar and spinor fields in arbitrary 
gravitational and electromagnetic backgrounds. 
This enables one to describe particles in a local, general-covariant 
and gauge-invariant way. However,
the current depends on the choice of a 2-point
function. There is a choice that leads to the local non-conservation
of the current in a gravitational or an electromagnetic background,
which describes local particle production consistent with the usual
global description based on the Bogoliubov transformation. 
The most natural choice based on the Green function 
calculated using the Schwinger-DeWitt method leads to
the local conservation of the current, provided that interactions 
with quantum fields  
are absent. Interactions with quantum fields lead to the local 
non-conservation of the current which describes local particle production
consistent with the usual
global description based on the interaction picture.  
\end{abstract}
\vspace*{0.9cm}

\section{Introduction}

One of the main problems regarding quantum field theory in curved spacetime
is how to introduce the concept of particles. 
The lack of a natural choice of the 
time coordinate implies that the generalization 
of the conventional definition of particles in Minkowski spacetime
is not unique in general spacetime \cite{ful,bd}. There are three main 
approaches to coping with this problem. Each of them has some 
advantages and disadvantages. Let us shortly describe them, emphasizing 
their disadvantages. 

The first approach is based on the theoretical 
point of view that the concept of particles
has no fundamental meaning in field theory \cite{davies}. 
Therefore, one should study only local, well-defined covariant 
operators, such as $T_{\mu\nu}(x)$, and try to express all 
observable quantities in terms of these. A disadvantage of such an 
approach is the fact that   
particles are what we observe in experiments. 
If we require that quantum field theory describes the observed 
objects, then it should describe particles. 

The second approach, complementary to the first one,   
resides on the physical point of view that particles are 
what particle detectors measure. Therefore, one avoids  
introducing a particle-number operator by studying the 
response of a model of a particle detector. These models are 
usually of the Unruh-DeWitt type \cite{unruh,dewitt2}. However, such an 
approach is unable to answer the following questions. 
If all observables in quantum mechanics are represented by 
hermitian operators and do not require a model of a corresponding 
detector, then why is the particle number an exception? 
Do Unruh-DeWitt type detectors describe the essential 
properties of real particle detectors (such as a Wilson chamber or 
a Geiger-M\"{u}ller counter) in real experiments? 

The third approach, perhaps the most popular one, is somewhere between 
the first two approaches. Although there may not exist a natural 
choice of the time coordinate in the whole spacetime, some regions 
of spacetime may have a natural choice of the time coordinate. 
Accordingly, one introduces different definitions of the particle-number  
operator for different regions. The two definitions are related by a 
non-trivial Bogoliubov transformation, which is interpreted as 
particle production \cite{bd,park,hawk} or as a variant of the Unruh effect
\cite{unruh,unruhwald,sciama}. In this way, 
one is able to talk about the average  
number of particles in some large region of spacetime. However, the average 
number of particles should make sense in any region of spacetime, 
especially if the wavelength of the particles is much smaller than the 
region. The third approach, based on the Bogoliubov transformation, 
is not local and general-covariant, so it  
is unable to answer questions such as the following. 
How to calculate, at least in principle, the average number of 
particles in an {\em arbitrary} region of spacetime? 
Is the average number of particles 
in the union of two non-intersecting
3-dimensional regions that lie on the same 3-dimensional spacelike
hypersurface equal to the sum of the average number of particles
in separate regions? If black holes 
radiate particles as predicted by Hawking \cite{hawk}, then what is the 
average particle distribution at, for example, 2 Schwarzschild radii 
away from the black-hole centre? 

There is a lot of similarity between the particle production 
by a background gravitational field and that by a background 
electromagnetic field \cite{step,paren,paren2}. In particular, 
just as the former has problems with general covariance, 
the description of the later through the Bogoliubov transformation 
has problems with gauge invariance \cite{padmprl}. This is related 
to the fact that gauge invariance is a local property, 
while the Bogoliubov-transformation method is based on a 
global definition of particles. 

Recently, a new approach to the concept of particles 
in quantum field theory
has been proposed \cite{nikol_PLB}. In this paper
we further develop the ideas introduced in \cite{nikol_PLB} and 
show that the new approach  
avoids the disadvantages of the other existing approaches
discussed above. 
This new approach is inspired by 
some advantages of the first and the third approach and does not 
rely on any model of a particle detector. The main idea is the 
assumption that there should exist a well-defined {\em local} 
operator of particle-number density. There are two reasons for 
such an assumption. First, this allows a general-covariant 
and gauge-invariant formulation. 
Second, although the particle-number operator is a 
global operator in the conventional approach,  
this assumption is consistent with the fact that 
particles appear as local objects in all existing experiments.

Our approach is based on a similarity
between the number of particles and charge. For complex fields,
the total number of particles is the sum of the number of particles and
antiparticles, while the total charge is the difference of these two
numbers. The concept of charge can be described in a local
and covariant manner because there exists a local vector current
of charge density. We find that a similar vector current exists
for the number of particles as well. Nevertheless, it appears that this
local current is not unique, but depends on the choice of a
2-point function. When a unique
(or a preferred) vacuum exists,
then the 2-point function is equal to the Wightman function
and the current is conserved.
When such a vacuum does not exist, then there is a
choice of the 2-point function that leads to the local non-conservation
of the current in a gravitational or electromagnetic background,
which describes local particle production consistent with the usual
global description based on the Bogoliubov transformation.
Another choice, based on the Green function
calculated using the Schwinger-DeWitt method \cite{schw,dewitt},  
seems to be more natural. This choice leads to
the local conservation of the current in an arbitrary 
gravitational or electromagnetic background,
provided that other interactions are absent.

Before studying the new local concept of particles
in Secs.~\ref{HERMIT}, \ref{SEC4}, and \ref{SEC5}, in Sec.~\ref{SEC2} we 
clarify some physical aspects of the concept of particles in 
the conventional approach. 
In Sec.~\ref{HERMIT} we study in detail the case of hermitian scalar fields. 
In Secs.~\ref{SEC4} and \ref{SEC5} we generalize the 
analysis to complex scalar fields and 
spin-1/2 fields, respectively. The conclusions and prospects for 
further investigations are given in Sec.~\ref{SEC6}. In the Appendix, 
we derive the particle-density operator in 
non-relativistic quantum field theory.

\section{Qualitative remarks on the concept of particles}
\label{SEC2}

In quantum field theory,
an $n$-particle state is defined as a normalized state of the form
\begin{equation}\label{n-part}
|n\rangle =\int d^3 k_1 \cdots d^3 k_n \,
f({\bf k}_1, \ldots ,{\bf k}_n)\, a^{\dagger}_{{\bf k}_1} \cdots
a^{\dagger}_{{\bf k}_n}|0\rangle  .
\end{equation}
The crucial question we attempt to answer in this
section is why such formally
defined states correspond to the states observed in
typical experiments as $n$ separated entities, i.e., ``particles".

We introduce the concept of a classical particle detector, such as
a Wilson chamber or a Geiger-M\"{u}ller counter. We call such
detectors classical, because, in order to understand how and why such
detectors respond, quantum field theory is not essential.
Classical detectors respond to states that correspond to the classical
concept of particle, i.e., to states which are well localized in
space. For example, if two localized particles
are very near each other, which can be achieved by a suitable choice
of the function $f({\bf k}_1,{\bf k}_2)$, then
a classical detector will see this state as one
particle. The response of a classical detector is also localized in 
time, in the sense that it can, with good precision, determine 
the instant of time at which the particle takes a certain position 
in space. In principle, the product of uncertainties 
$\Delta E\Delta t$ may be arbitrarily small \cite{ahar}. 
Because of these local features, one does not expect that 
the response of a classical detector depends on global 
aspects such as the whole detector's trajectory or the existence of 
an event  
horizon. Therefore, classical detectors are very different 
from particle detectors of the Unruh-DeWitt type. 
Because of locality, one expects that 
classical particle detectors detect objects 
that are covariant with respect to general coordinate transformations. 
Therefore, one needs a local and general-covariant notion of a 
particle in quantum field theory.  

As we mentioned above, a many-particle state can be detected as 
one particle if these localized particles are very near each other. 
However, it seems that such a situation does not occur in practice. Why?
Before answering this question, first consider how a one-particle
state is detected.
Assume that a particle detector is localized in space. The function
$f({\bf k}_1)$ may correspond to a plane wave or to a state with two lumps,
but such states cannot be detected by the localized detector. 
Such a state will not be detected until the
function $f({\bf k}_1)$ collapses to a one-lump state. (We do not attempt to
answer the question whether the wave function collapses
spontaneously or the detector causes the collapse. In some interpretations of
quantum mechanics there is a clear answer to this question, but we
do not want to prefer any particular interpretation.)

Now consider a two-particle state. Assume that, {\it a priori}, all
functions $f({\bf k}_1,{\bf k}_2)$ are equally probable. 
(This probability does not reflect a stochastic nature of quantum states, 
but simply our ignorance of the details that govern the dynamics and the 
initial conditions.)
However,
the number of functions corresponding  to a two-lump state is
much larger than that for a one-lump state,
so it is very improbable that two particles will form
one lump. (Of course, the numbers of two-lump states and one-lump states
are both infinite, but their ratio is not equal to 1.
To visualize this, note that if there are $N$ places where a lump can be
placed, then (for a fixed number of particles) 
there are $N$ different one-lump states and 
$N^2-N$ different two-lump states.) 
A four-lump state,
for instance, is even more probable
than a two-lump state, but if one particle is distributed
in more than one lump, it will not be detected.

In this way we have explained why an $n$-particle
state behaves as an $n$-lump state in
experiments;
a particle splited in more than one lump will not be detected, while a 
smaller number of lumps can be realized in a smaller number of ways, so it 
has a smaller probability.
Of course, if there are strong attractive forces among
fields, there may exist a natural tendency for $n$-particle states to form
a one-lump state (hadrons, $\alpha$-particles), in which case it is more
convenient to treat such states as one-particle states. Actually, it is
often completely incorrect to treat such states as $n$-particle states,
because non-perturbative effects of interactions may completely
change the spectrum of states in a free theory, such as is the case
for QCD.

One may argue that classical detectors are not the best operational
way to define a particle. One should rather study quantum detectors.
(Even the response of a ``classical" detector should be
ultimately described by quantum field theory.)
For example, an atom may be viewed as a quantum photon detector 
as it absorbs precisely one photon, not a half of it, nor
two photons. It is often thought that this is an inherent 
quantum property.  
However, it is important to emphasize that there is not
any deep, fundamental principle that forbids absorption or emission
of a half of a particle or two particles. 
It is merely a consequence of a particular
form of
dynamics. For example, an electron in the atom absorbs {\em one} photon
because
the interaction Lagrangian 
${\cal L}_{{\rm int}} =e\bar{\psi}\gamma_{\mu}\psi A^{\mu}$
is {\em linear} in $A^{\mu}$. Of course, higher-order corrections allow
absorption of two photons (sum of their energies must be equal to the
difference of energies of the atom levels), but such processes are
suppressed dynamically (small coupling constant) and kinematically (small
probability of a one-lump two-photon state). However, is it, in principle,
possible to absorb a half of a particle? One can exclude such a
possibility by proposing that the Lagrangian ${\cal L}(\phi,
\partial_{\mu}\phi, \ldots )$
of fundamental fields
should be an analytic function around zero. For example,
assuming that there are no derivative couplings, the analyticity implies
that a local interaction Lagrangian has a form
\begin{equation}\label{lagr}
{\cal L}_{{\rm int}}(\phi(x),\cdots ) =\sum_{n\geq 0}C_n(x)(\phi(x))^n  ,
\end{equation}
where $n$ are integers and $C_n(x)$ depend on some other fields.
The interaction (\ref{lagr})
implies that the number of absorbed or emitted particles must be
an integer. If $\phi(x)$ is an effective, composite field, like
$\phi(x)=\chi(x)\chi(x)$, then one can have a term
proportional to $\sqrt{\phi}=\chi$, in which case a ``half" of the
$\phi$-particle, i.e., one $\chi$-particle, can be absorbed or emitted.
If one allows terms like $\sqrt{\phi}$ for fundamental fields,
then even a ``half" of a fundamental particle can be absorbed or emitted.
(In this case, the concept of particle based on perturbative
calculations
is no longer a good concept. In particular, it is not clear,
even algebraically, what $\sqrt{a^{\dagger}}|0\rangle$ is.)

\section{Hermitian scalar field}\label{HERMIT}

The hermitian scalar field is the simplest example of a quantum field. 
Therefore, we start the quantitative analysis by studying 
this simplest case in detail. The first two subsections do not contain 
new results, but they are written for the sake of completeness and easier 
understanding of our later results.

\subsection{Canonical quantization}

Let $g_{\mu\nu}$ be a classical background metric, $g$ the determinant of 
$g_{\mu\nu}$, and $R$ the curvature calculated from $g_{\mu\nu}$. 
The action of a hermitian scalar field $\phi$ is 
\begin{equation}\label{1}
S=\int d^4 x \, |g|^{1/2} \,\frac{1}{2} [g^{\mu\nu}(\partial_{\mu}\phi)
(\partial_{\nu}\phi)-m^2 \phi^2 -\xi R\phi^2]  ,
\end{equation}
where $\xi$ is a coupling constant.
Writing this as 
\begin{equation}\label{tildaL}
S=\int d^4x \, |g|^{1/2}{\cal L}, 
\end{equation}
the canonical-momentum vector is
\begin{equation}\label{2}
\pi_{\mu} =\frac{\partial {\cal L}}{\partial 
(\partial^{\mu}\phi)} =\partial_{\mu}\phi  .
\end{equation}
The correponding equation of motion is 
\begin{equation}\label{8}
(\nabla^{\mu}\partial_{\mu}+m^2 +\xi R)\phi=0  ,
\end{equation}
where $\nabla^{\mu}$ is the covariant derivative.
Let $\Sigma$ be a spacelike Cauchy hypersurface with a unit vector $n^{\mu}$ 
normal to $\Sigma$. The canonical-momentum scalar is defined as 
\begin{equation}
\pi =n^{\mu}\pi_{\mu}  .
\end{equation}
The volume element on $\Sigma$ is
\begin{equation}\label{volume_element}
d\Sigma^{\mu}=d^3x \, |g^{(3)}|^{1/2} n^{\mu}  .
\end{equation}
The scalar product is defined as
\begin{equation}\label{9}
(\phi_1,\phi_2)=i\cint^{\mu}\phi_1^* \apm \phi_2  ,
\end{equation}
where $a\apm b=a\partial_{\mu}b - (\partial_{\mu}a) b$. 
If $\phi_1$ and $\phi_2$ are solutions of (\ref{8}), then (\ref{9})
does not depend on $\Sigma$ \cite{ford}.

We choose coordinates $(t,{\bf x})$ such that $t=$constant
on $\Sigma$. In these coordinates 
\begin{equation}
n^{\mu}=g^{\mu}_{0}/\sqrt{g_{00}}  ,
\end{equation}
and the canonical commutation relations are
\begin{equation}\label{3} 
[\phi(x),\phi(x')]_{\Sigma}=[\pi(x),\pi(x')]_{\Sigma}=0  ,
\end{equation}
\begin{equation}\label{4}
[\phi(x),\pi(x')]_{\Sigma}=|g^{(3)}|^{-1/2}\, i\delta^3 ({\bf x}-{\bf x}')  .
\end{equation}
The label ${\Sigma}$ denotes that $x$ and $x'$ lie on ${\Sigma}$.
Eq.~({\ref{4}) can be written in a manifestly covariant form as 
\begin{equation}\label{5}
\cint'^{\mu}[\phi(x),\partial'_{\mu}\phi(x')]\chi(x')=
\cint'^{\mu}[\phi(x'),\partial_{\mu}\phi(x)]\chi(x')=i\chi(x)  ,
\end{equation}
where $\chi(x')$ is an arbitrary test function. 

For practical calculations, it is more convenient to introduce 
the quantity $\tilde{n}^{\mu}$ defined as
\begin{equation}
\tilde{n}^{\mu}=|g^{(3)}|^{1/2}n^{\mu}  ,
\end{equation}   
where the tilde on $\tilde{n}^{\mu}$ denotes
that it is not a vector. It has the property 
$\nabla_{\mu} \tilde{n}_{\nu}=0$ and has the explicit form
\begin{equation}\label{7.2}
\tilde{n}^{\mu}=(|g^{(3)}|^{1/2}/\sqrt{g_{00}},0,0,0)  .
\end{equation}
The volume element (\ref{volume_element}) can be written as 
\begin{equation}\label{7.1}
d\Sigma^{\mu}=d^3x \, \tilde{n}^{\mu}  ,
\end{equation}
while (\ref{4}) can be written as 
\begin{equation}\label{6}
\tilde{n}^0(x') [\phi(x),\partial'_0\phi(x')]_{\Sigma} 
=i \delta^3({\bf x}-{\bf x}') .
\end{equation} 
This can also be written in the canonical form as
\begin{equation}\label{canon_pi}
[\phi(x),\tilde{\pi}(x')]_{\Sigma}=i \delta^3({\bf x}-{\bf x}'),
\end{equation}
where
\begin{equation}
\tilde{\pi}=|g^{(3)}|^{1/2}\pi
\end{equation}
is not a scalar.

\subsection{A choice of representation}

Let us choose a particular complete orthonormal set of solutions 
$\{ f_k(x) \}$ of Eq.~(\ref{8}). They satisfy
\begin{eqnarray}\label{11}
& (f_k,f_{k'})=-(f^*_k,f^*_{k'})=\delta_{kk'} , & \nonumber \\
& (f^*_k,f_{k'})=(f_k,f^*_{k'})=0  . &
\end{eqnarray}
The field $\phi$ can be expanded as
\begin{equation}\label{10}
\phi(x)=\sum_{k}a_k f_k(x)+ a^{\dagger}_k f^*_k(x)  .
\end{equation}
From (\ref{11}) and (\ref{10}) we find
\begin{equation}\label{13}
a_k=(f_k,\phi)  , \;\;\; a^{\dagger}_k=-(f^*_k,\phi),
\end{equation}
while from (\ref{11}) and the canonical commutation relations 
it follows that
\begin{eqnarray}\label{12}
& [a_k,a_{k'}^{\dagger}]=\delta_{kk'}  , & \nonumber \\
& [a_k,a_{k'}]=[a_k^{\dagger},a_{k'}^{\dagger}]=0  . &
\end{eqnarray}
Therefore, we can interpret $a_k$ and $a_k^{\dagger}$ as lowering 
and raising operators, respectively. They induce the representation 
of the field algebra in the usual way. The vacuum $|0\rangle$ 
of the corresponding Hilbert space is defined by
\begin{equation}\label{vacuum}
a_k|0\rangle=0  .
\end{equation} 
The operator of the total number of particles is
\begin{equation}\label{14}
N=\sum_{k}a^{\dagger}_k a_k  .
\end{equation}

We also introduce the function $W(x,x')$, here defined as
\begin{equation}\label{18r}
W(x,x')=\sum_{k} f_k(x) f^*_k(x')  .
\end{equation}
(Later we also study different definitions of $W(x,x')$.)
Using (\ref{10}), we find that it is a Wightman function
\begin{equation}\label{18l}
W(x,x')=\langle 0|\phi(x)\phi(x') |0\rangle  .
\end{equation}
From (\ref{18r}) it follows that 
\begin{equation}\label{22}
W^*(x,x')=W(x',x)  .
\end{equation}
From the fact
that $f_k$ and $f^*_k$ satisfy Eq.~(\ref{8}) we find
\begin{equation}\label{21}
(\nabla^{\mu}\partial_{\mu}+m^2 +\xi R(x))W(x,x')=0  ,
\end{equation}
\begin{equation}\label{21'}
(\nabla'^{\mu}\partial'_{\mu}+m^2 +\xi R(x'))W(x,x')=0  .
\end{equation}
From (\ref{18l}) and the canonical commutation relations we find
that $f_k$ and $f^*_k$ are functions such that
\begin{equation}\label{19}
W(x,x')|_{\Sigma}=W(x',x)|_{\Sigma}  ,
\end{equation}
\begin{equation}\label{19'}
\partial_0\partial'_0 W(x,x')|_{\Sigma}=
\partial_0\partial'_0 W(x',x)|_{\Sigma}  ,
\end{equation}
\begin{equation}\label{20}
\tilde{n}^0\partial'_0[W(x,x')-W(x',x)]_{\Sigma}=
i\delta^3({\bf x}-{\bf x}')  .
\end{equation}

\subsection{The current of particle density}

The number operator given by (\ref{14}) is a global quantity. 
However, as originally found in \cite{nikol_PLB}, a new way of looking into
the concept of particles emerges when (\ref{13}) is put into
(\ref{14}), and (\ref{9}) and (\ref{18r}) are used. This leads to 
\begin{equation}
N = \cint^{\mu} \cint'^{\nu} W(x,x')\apm \; \apn \phi(x)\phi(x')  .
\end{equation}
By interchanging the names of the coordinates $x$ and $x'$ and the names 
of the indices $\mu$ and $\nu$, this can be  
written as a sum of two equal terms
\begin{eqnarray}\label{N=sum}
N & = & \frac{1}{2} \cint^{\mu} \cint'^{\nu} W(x,x')\apm \; 
 \apn \phi(x)\phi(x') \nonumber \\ 
 & + & \frac{1}{2} \cint^{\mu} \cint'^{\nu} W(x',x)\apm \;
 \apn \phi(x')\phi(x)  .
\end{eqnarray}
Using also (\ref{22}), we see that (\ref{N=sum}) 
can be written in the final form as 
\begin{equation}\label{15}
N=\cint^{\mu} j_{\mu}(x)  ,
\end{equation}
where the vector $j_{\mu}(x)$ is the hermitian operator 
\begin{equation}\label{16}
j_{\mu}(x)=\cint'^{\nu} \frac{1}{2} \{W(x,x')\apm \; \apn
\phi(x)\phi(x') + {\rm h.c.} \}.
\end{equation}
Obviously, the vector $j_{\mu}(x)$ should be interpreted as the
local current of particle density. The representation of $N$ by 
(\ref{15}) and (\ref{16}) has three advantages with respect to the 
conventional representation (\ref{14}). First, it avoids the use 
of the operators $a_k$ and $a_k^{\dagger}$ related to a particular 
choice of the modes $f_k(x)$. Second, it is manifestly covariant. 
Third, the local current $j_{\mu}(x)$ allows to view the concept of 
particles in a local manner, which the conventional 
representation does not allow.

If we put (\ref{10}) and (\ref{18r}) in (\ref{16}) and then use 
(\ref{9}), (\ref{11}), (\ref{12}), and the antisymmetry of the 
operator $\apm \,$, we find
\begin{equation}\label{22_mu}
j_{\mu}=i\sum_{k,k'}f_k^* \apm f_{k'} a_k^{\dagger}a_{k'}  .
\end{equation}
From this form we see that $j_{\mu}$ is automatically normally 
ordered and has the property
\begin{equation}
j_{\mu}|0\rangle =0  .
\end{equation}
This is actually not surprising because we have started from the 
operator (\ref{14}) which is also normally ordered. 

Using (\ref{8}) and (\ref{21}), we find that
the current (\ref{16}) possesses the property
\begin{equation}\label{24}
\nabla^{\mu}j_{\mu}(x)=0  .
\end{equation}
This covariant conservation law means
that the background gravitational field does not produce particles,
provided that a unique (or a preferred) vacuum defined by 
(\ref{vacuum}) exists. 
(This local covariant conservation implies also the global conservation
because it provides that (\ref{15}) does not depend on time. The extra terms 
in (\ref{24}) that originate from the fact that 
$\nabla_{\mu}\neq\partial_{\mu}$ are compensated by the extra terms in 
(\ref{15}) that originate from the fact that 
$d\Sigma^{\mu}\neq (d^3x,0,0,0)$.)
The choice of the vacuum is related 
to the choice of the function $W(x,x')$ (see (\ref{18r}) and 
(\ref{18l})).
 
Note that although $j_{\mu}(x)$ is a local operator, some non-local
features of the particle concept still remain, because (\ref{16})
involves an integration over $\Sigma$ on which $x$ lies. Since
$\phi(x')$ satisfies (\ref{8}) and $W(x,x')$ satisfies
(\ref{21'}), this integral does not depend on $\Sigma$. However, 
it does depend on the choice of $W(x,x')$. Note also that
the separation between $x$ and $x'$ in (\ref{16}) is spacelike,   
which softens the non-local features because $W(x,x')$
decreases rapidly with spacelike separation. As can be explicitly
seen with the usual plane-wave modes in Minkowski spacetime 
(see Sec.~\ref{MINK}),
$W(x,x')$ is negligible when the spacelike separation is much larger
than the Compton wavelength $m^{-1}$.

\subsection{The case of Minkowski spacetime}\label{MINK}

In Minkowski spacetime with the metric
$g_{\mu\nu}={\rm diag}(1,-1,-1,-1)$, it is  
most natural to choose the modes 
$f_k$ as the usual plane-wave modes $f_{{\bf k}}$, where ${\bf k}$ 
is the 3-momentum. Using the discrete notation for the summation 
over ${\bf k}$, these modes are
\begin{equation}
f_{{\bf k}}(x)=\frac{e^{-ikx}}{\sqrt{V2\omega_{{\bf k}}}}  ,
\end{equation}      
where $\omega_{{\bf k}}=\sqrt{{{\bf k}}^2+m^2}$, $V$ is the 3-volume and
$k^{\mu}=(\omega_{{\bf k}},{\bf k})$.
Eq.~(\ref{22_mu}) becomes 
\begin{equation}\label{22mink}
j_{\mu}=\frac{1}{V}\sum_{{\bf k},{\bf k}'}
\frac{k_{\mu}+k'_{\mu}}{2\sqrt{\omega_{{\bf k}}\omega_{{\bf k}'}}} 
\, e^{i(k-k')x}\, a_{{\bf k}}^{\dagger}a_{{\bf k}'} .
\end{equation}
The physical content is more obvious from the space-integrated operator
\begin{equation}\label{23}
\tilde{J}_{\mu}\equiv \int d^3x \, j_{\mu}(x)=\sum_{{\bf k}}
\frac{k_{\mu}}{\omega_{{\bf k}}} \, a^{\dagger}_{{\bf k}}
a_{{\bf k}}  ,
\end{equation}
where the tilde on $\tilde{J}_{\mu}$ denotes that it does not 
transform as a 4-vector.
The quantity $k^i/\omega_{{\bf k}}$ is the 3-velocity $v^i$, so
we find  
\begin{equation}\label{23'}
\tilde{{\bf J}}|n_q\rangle ={\bf v}n_q |n_q\rangle  ,
\end{equation}
where $|n_q\rangle =(n_q!)^{-1/2} (a_q^{\dagger})^{n_q} |0\rangle$ 
is the state with $n_q$
particles with the momentum $q$. Eqs.~(\ref{23}) and (\ref{23'})
support the interpretation of $j_{\mu}$ as the particle current.
 
The two-point function (\ref{18r}) is 
\begin{equation}\label{prop_diskont}
W(x,x')=\frac{1}{V} \sum_{{\bf k}} \frac{e^{-ik(x-x')}}{2\omega_{{\bf k}}} .
\end{equation}
In the infinite-volume limit, we perform the replacement
\begin{equation}
\sum_{{\bf k}} \rightarrow V\int \frac{d^3k}{(2\pi)^3} ,
\end{equation}
so (\ref{prop_diskont}) takes the usual form
\begin{equation}\label{prop_kont}
W(x,x')=\int \frac{d^3k}{(2\pi)^3 2\omega_{{\bf k}}} 
\, e^{-ik(x-x')}  .
\end{equation}
The properties of this function, often denoted by $\Delta(x-x')$ \cite{bd2}, 
are well known. In particular, for $t=t'$ and 
$|{\bf x}-{\bf x}'|\equiv r$, it behaves asymptotically as
\begin{equation}\label{asimp1}
\Delta(r) \propto \frac{\sqrt{mr}}{r^2} e^{-mr} \;\; {\rm for}
\;\; rm\gg 1  . 
\end{equation}
For $m=0$, the function $W(x,x')$ is
\begin{equation}\label{asimp2}
\Delta(x-x')|_{m=0}=-\frac{1}{4\pi^2}\left[ \frac{1}{(x-x')^2}
+i\pi \delta((x-x')^2) \right]  .
\end{equation}
The function (\ref{prop_kont}) is also equal to $G^+(x,x')$, 
where $iG^+(x,x')$ is equal to the Green function
\begin{equation}
G(x,x')=\int \frac{d^4k}{(2\pi)^4}
\frac{e^{-ik(x-x')}}{k^2-m^2}  ,
\end{equation}
calculated with the appropriately chosen contour of integration 
over $k_0$ \cite{bd}. 

\subsection{The question of positivity of the local number of particles}

Eq.~(\ref{15}) can be written as 
\begin{equation}
N=\int_{\Sigma} d^3x \, |g^{(3)}|^{1/2} n(x)  ,
\end{equation}
where 
\begin{equation}
n=n^{\mu}j_{\mu} .
\end{equation}
Since the total number of particles (\ref{14}) is a non-negative operator 
(in the sense that the expected value $\langle\psi|N|\psi\rangle$ 
is non-negative for any state $|\psi\rangle$), one could expect that 
the local density of particles $n(x)$ is also a 
non-negative operator. For example, in the Minkowski case, for 
$|q\rangle \equiv a_{{\bf q}}^{\dagger}|0\rangle$, we find 
$\langle q|n|q \rangle =\langle q|j_0|q \rangle =V^{-1}$, which is a
positive constant. However, for $|\psi\rangle =2^{-1/2}(|q_1\rangle +
|q_2\rangle)$, where $q_1\neq q_2$, we find
\begin{equation}\label{number53}
\langle\psi|j_0|\psi\rangle =\frac{1}{V} \left[ 1+
\frac{\omega_{{\bf q}_1}+\omega_{{\bf q}_2}}{2\sqrt{\omega_{{\bf q}_1}
\omega_{{\bf q}_2}}} \, {\rm cos}(q_1-q_2)x \right]  ,
\end{equation}
which is negative for some values of $x$, provided that $\omega_{{\bf q}_1}
\neq \omega_{{\bf q}_2}$. Therefore, the particle density $n(x)$ is not 
a non-negative operator. Is this in contradiction with the experimental 
fact that the number of particles in a small volume $\sigma\subset\Sigma$ 
cannot be negative? The answer is {\em no}! When the position of a particle 
is measured with some accuracy such that it is (almost) certain that the
particle is localized inside $\sigma$, then the state $|\psi\rangle$ 
of the particle 
is a localized state, such that the corresponding wave function  
vanishes (or is negligible) outside of $\sigma$. 
Consequently, the quantity $\langle\psi|n(x)|\psi\rangle$ 
vanishes (or is negligible)
outside of $\sigma$. Therefore, the total number of particles 
in this state is (approximately) equal to
\begin{equation}\label{Nsig=}
{\cal N}_{\sigma}=\int_{\sigma} d^3x \, |g^{(3)}|^{1/2} 
\langle\psi|n(x)|\psi\rangle .
\end{equation}
Since the total number of particles cannot be negative, it follows that the 
number of particles inside $\sigma$ is non-negative for such a localized 
state. 

Let us illustrate the heuristic arguments above by a concrete example. 
We study the case of Minkowski spacetime with infinite volume $V$, so the 
field expansion (\ref{10}) becomes
\begin{equation}
\phi(x)=\int d^3k [a_{{\bf k}}f_{{\bf k}}(x)+
a_{{\bf k}}^{\dagger}f^*_{{\bf k}}(x)]  ,
\end{equation}
where
\begin{equation}
f_{{\bf k}}(x)=\frac{e^{-ikx}}{\sqrt{(2\pi)^3 2\omega_{{\bf k}} }}  .
\end{equation}
Let the state of the system be the one-particle state
\begin{equation}\label{state_y}
|{\bf y}\rangle \equiv c^{-1/2} \phi(0,{\bf y}) |0\rangle .
\end{equation}
The constant $c$ is the norm determined by the requirement that 
$\langle {\bf y}|{\bf y}\rangle =1$, so it is equal to
\begin{equation}
c=\Delta^+(0)=\Delta^-(0)  ,
\end{equation}
where
\begin{equation}
\Delta^{\pm}(x)=\int \frac{d^3k}{(2\pi)^3 2\omega_{{\bf k}}} 
e^{\mp ikx}  .
\end{equation}
The wave function corresponding to the state (\ref{state_y}) is
\begin{equation}\label{wf_y}
\psi_{{\bf y}}(x)=\tilde{c}\langle 0|\phi(x)|{\bf y}\rangle =
\tilde{c}c^{-1/2}\Delta^+(t,{\bf x}-{\bf y})  ,
\end{equation}
where $\tilde{c}$ is a constant chosen such that the wave function 
is normalized in an appropriate way. This wave function is related 
to a Lorentz-invariant notion of particle detection \cite{rovel}.
Using the fact that $W(x,x')=\Delta^+(x-x')$, a straightforward calculation 
gives
\begin{equation}
\langle {\bf y}|j_0(x)|{\bf y}\rangle =\frac{1}{2}
\left[ \frac{ \Delta^+(t,{\bf x}-{\bf y}) \delta^-(t,{\bf x}-{\bf y})}
{\Delta^+(0)} +{\rm c.c.} \right] , 
\end{equation} 
where
\begin{equation}
\delta^{\pm}(x)=\int \frac{d^3k}{(2\pi)^3} e^{\mp ikx}  .
\end{equation}
In particular, at $t=0$
\begin{equation}
\langle {\bf y}|j_0(0,{\bf x})|{\bf y}\rangle =\delta^3 ({\bf x}-{\bf y}) ,
\end{equation}
which implies that the state (\ref{state_y}) has a non-negative and 
strictly localized particle density at $t=0$. Note that the wave function 
(\ref{wf_y}) is not strictly localized at $t=0$, but decreases rapidly 
with $|{\bf x}-{\bf y}|$ (see (\ref{asimp1}) and (\ref{asimp2})).

Note also that in a typical physical situation in which particles are 
detected (see Sec.~\ref{SEC2}), the state $|\psi\rangle$ corresponds 
to $N$ particles, where each particle is localized inside one of 
$N$ lumps. In such a situation, (\ref{Nsig=}) represents the number 
of particles in one of the lumps, so this number is equal to one.  
 
We also stress that negative particle densities do not appear in 
non-relativistic quantum field theory (see Appendix A). 
Heuristically, this can be seen from the frequency-dependent factor 
that multiplies the cosine function in (\ref{number53}). Since in the 
non-relativistic limit $\omega_q\simeq m$, this factor 
becomes equal to 1, leading to a non-negative density.

Note finally that, in general, the particle density (that may be negative) 
is not the same thing as the probability density (that cannot be negative) 
of finding the particle at a position ${\bf x}$. The probability density 
in a state $|\psi\rangle$ is equal to 
$|\langle\tilde{{\bf x}}|\psi\rangle|^2$, where $|\tilde{{\bf x}}\rangle$ 
is an eigenstate of the position operator ${\bf X}$ with the eigenvalue 
${\bf x}$. For the relativistic case, the problem of finding the operator 
${\bf X}$ and the states $|\tilde{{\bf x}}\rangle$ is not yet satisfactorily 
solved. In the nonrelativistic limit, the states $|\tilde{{\bf x}}\rangle$ 
are equal to the states $|{\bf x}\rangle$ defined by (\ref{state_y}).

\subsection{Particle production by non-gravitational interactions}

Let us now study the case in which a non-gravitational interaction 
represented by ${\cal L}_{{\rm int}}=-{\cal H}_{{\rm int}}$
is also present. In this case, the equation of motion is
\begin{equation}\label{35}   
(\nabla^{\mu}\partial_{\mu}+m^2 +\xi R)\phi=J ,
\end{equation}
where $J(x)$ is a local operator containing $\phi$
and/or other dynamical quantum fields. Since it describes the interaction,
it does not contain terms linear in quantum fields.
(For example, when ${\cal L}_{{\rm int}}=-\lambda\phi^4/4$, then
$J(x)=-\lambda\phi^3(x)$).
We propose that even in this general case
the particle current is given by (\ref{16})
\begin{equation}\label{16'}
j_{\mu}(x)=\cint'^{\nu} \frac{1}{2} \{W(x,x')\apm \; \apn
\phi(x)\phi(x') + {\rm h.c.} \},
\end{equation}
where $W(x,x')$ is the same function
as before, satisfying the ``free" equations
\begin{equation}\label{free_W}
(\nabla^{\mu}\partial_{\mu}+m^2 +\xi R(x))W(x,x')=0  ,
\end{equation}
\begin{equation}\label{free_W'}
(\nabla'^{\mu}\partial'_{\mu}+m^2 +\xi R(x'))W(x,x')=0  ,
\end{equation}
and having the expansion
\begin{equation}\label{free_exp}
W(x,x')=\sum_{k} f_k(x) f^*_k(x') .
\end{equation}
As we show below, such an ansatz leads to particle production
consistent with the conventional approach to
particle production caused by a non-gravitational interaction 
${\cal L}_{{\rm int}}$.

The field $\phi$ in (\ref{16'}) satisfies (\ref{35}). Therefore, 
using (\ref{free_W}), we find
\begin{equation}\label{36}
\nabla^{\mu}j_{\mu}(x) = \cint'^{\nu} \frac{1}{2}\{
W(x,x')\apn J(x)\phi(x') +{\rm h.c.}\}  .
\end{equation}
Note that only $x$ (not $x'$) appears
as the argument of $J$ on
the right-hand side of (\ref{36}), implying that $J$ plays a strictly
local role in particle production.

Let us now show that our covariant description (\ref{36}) of particle
production is consistent with the conventional approach to
particle production caused by a non-gravitational interaction.
Let $\Sigma(t)$ denote some foliation of spacetime into
Cauchy spacelike hypersurfaces. The total mean number of particles
at the time $t$ in a state $|\psi\rangle$ is
\begin{equation}\label{37}
{\cal N}(t)=\langle \psi |N(t)|\psi\rangle  ,
\end{equation}
where
\begin{equation}\label{38}
N(t)=\int_{\Sigma(t)} d\Sigma^{\mu} j_{\mu}(t,{\bf x})  .
\end{equation}
Equation (\ref{37}) is written in the Heisenberg picture. However,
matrix elements do not depend on picture. We introduce the
interaction picture, where the interaction Hamiltonian is the
part of the Hamiltonian that generates the right-hand side of
(\ref{35}). The state $|\psi\rangle$ transforms to a time-dependent
state 
\begin{equation}
|\psi_{{\rm int}}(t)\rangle =U_{{\rm int}}(t)|\psi\rangle  .
\end{equation}
Here the unitary operator $U_{{\rm int}}(t)$ satisfies 
\begin{equation}\label{dUdt}
i\frac{d}{dt}U_{{\rm int}}(t)=\tilde{H}_{{\rm int}}(t)U_{{\rm int}}(t)  ,
\end{equation}
where
\begin{equation}\label{Hint}
\tilde{H}_{{\rm int}}(t)=
\int_{\Sigma(t)} d^3x \, |g|^{1/2} {\cal H}_{{\rm int}}(x).
\end{equation}
Note that Eqs.~(\ref{dUdt}) and (\ref{Hint}) are not manifestly 
covariant. The covariance of the 
time-evolution law is discussed in Sec.~\ref{COVAR}.
In the interaction picture, the field $\phi$ satisfies the free
equation (\ref{8}), so the expansion (\ref{10}) can be used. 
This, together with (\ref{38}), (\ref{16'}), and (\ref{free_exp}),   
implies that $N(t)$ transforms to
\begin{equation}
N_{{\rm int}}=\sum_{k}a^{\dagger}_k a_k.
\end{equation}
Therefore, (\ref{37}) can be written as
\begin{equation}\label{39}
{\cal N}(t)=\langle \psi_{{\rm int}}(t)| N_{{\rm int}} 
|\psi_{{\rm int}}(t)\rangle=
\sum_{k}\langle\psi_{{\rm int}}(t)|a^{\dagger}_k a_k
|\psi_{{\rm int}}(t)\rangle  ,
\end{equation}
which is the usual formula that describes particle production caused by
a quantum non-gravitational interaction. This is most easily seen when 
$t\rightarrow\infty$ and the solution of (\ref{dUdt}) is chosen such 
that $|\psi_{{\rm int}}(t\rightarrow-\infty)\rangle =|\psi\rangle$, 
in which case $U_{{\rm int}}$ is the $S$-matrix operator. 

\subsection{Particle production by the gravitational background}
\label{PPGB}

So far, our discussion has been based on a particular choice 
of the modes $\{ f_k(x) \}$ as preferred modes. In practice, 
they are chosen such that $f_k$ are positive frequency solutions 
of (\ref{8}). However, when metric is time dependent, then 
such preferred modes do not exist. In the
conventional approach to the concept of particles, this is 
related to particle production by the gravitational background 
\cite{bd,park}. 
In this case, 
one introduces a new set of functions $u_l(x)$ for
each time $t$, such that $u_l(x)$ are positive-frequency modes
at that time.
This means that the modes $u_l$
possess an extra time dependence, i.e., they become
functions of the form $u_l(x;t)$. These functions do not
satisfy (\ref{8}). However, the functions $u_l(x;\tau)$
satisfy (\ref{8}), provided that $\tau$ is kept fixed
when the derivative $\partial_{\mu}$ acts on $u_l$. 
To describe the local particle production, we take
\begin{equation}\label{Wcreate}
W(x,x')=\sum_l u_l(x;t)u^*_l(x';t') ,
\end{equation}
instead of (\ref{18r}). Since $u_l(x;t)$ do not satisfy (\ref{8}),
the function (\ref{Wcreate}) does not satisfy (\ref{21}).
Instead, we have
\begin{equation}\label{m2}
(\nabla^{\mu}\partial_{\mu}+m^2 +\xi R(x))W(x,x')
\equiv -K(x,x') \neq 0 .
\end{equation}
Using (\ref{8}) and (\ref{m2}) in (\ref{16}), we find a relation similar
to (\ref{36}):
\begin{equation}\label{m3}
\nabla^{\mu}j_{\mu}(x)=\cint'^{\nu} \frac{1}{2}\{
K(x,x')\apn \phi(x)\phi(x')+{\rm h.c.} \}  .
\end{equation}

This local description of particle production is
consistent with the usual
global description based on the Bogoliubov transformation.
This is because (\ref{38}) and (\ref{16}) with (\ref{Wcreate})
and (\ref{10}) lead to
\begin{equation}\label{N(t)}
N(t)=\sum_l A^{\dagger}_l(t) A_l(t)  ,
\end{equation}
where
\begin{equation}\label{e10}
A_l(t)=\sum_{k} \alpha^*_{lk}(t) a_k - \beta^*_{lk}(t) a^{\dagger}_k  ,
\end{equation}
\begin{equation}\label{e7}
\alpha_{lk}(t)=(f_k,u_l)  , \;\;\;\; \beta_{lk}(t)=-(f^*_k,u_l)  .
\end{equation}
The time dependence of the Bogoliubov coefficients $\alpha_{lk}(t)$ and
$\beta_{lk}(t)$ is related to the extra time dependence of
the modes $u_l(x;t)$. If we assume that the
change of the average number of particles is slow, i.e., that
\begin{equation}
\partial_t A_l(t)\approx 0,
\end{equation}
which occurs when
\begin{equation}
\partial_t u_l(\tau,{\bf x};t)|_{\tau=t}\approx 0, 
\end{equation}
then the Bogoliubov
coefficients (\ref{e7}) are equal to the usual Bogoliubov
coefficients. This approximation is nothing else but the
adiabatic approximation,
which is a usual part of the convential
description of particle production \cite{park}.

\subsection{The natural choice of $W(x,x')$}\label{NATURAL}

In general,
there is no universal natural choice for the modes $u_l(x;t)$.
In particular, in a given spacetime, the choice of the natural
modes $u_l(x;t)$ may depend on the observer. If different
observers (that use different coordinates) use different modes for the
choice of (\ref{Wcreate}), then
the coordinate transformation alone does not describe how
the particle current is seen by different observers. In this sense,
the particle current is not really covariant. There are also
other problems related to the case in which different
observers use different modes \cite{nikolmpl}.

Since covariance was our original aim, it is desirable to
find a universal natural choice of $W(x,x')$, such that,
in Minkowski spacetime, it
reduces to the usual plane-wave expansion (\ref{prop_kont}).
Such a choice exists. This is based on the fact 
that the Feynman Green function
$G_F(x,x')$ can be calculated using the Schwinger-DeWitt method 
and that the knowledge of the Feynman Green function automatically yields 
the knowledge of other Green functions \cite{bd,dewitt}, such as 
$G^+(x,x')$. The Schwinger-DeWitt method does not require a choice of 
a particular set of modes.
Therefore, we propose
\begin{equation}\label{w=g}
W(x,x')=G^+(x,x'),
\end{equation}
where $G^+$ is calculated from $G_F$ given by
\begin{equation}\label{integral_s}
G_F(x,x')=i[g(x)g(x')]^{-1/4}\int_0^{\infty} ds\langle x,s|x',0\rangle .
\end{equation}
The function $\langle x,s|x',0\rangle$ is determined by 
the function $\sigma(x,x')$ and its derivatives, where $\sigma(x,x')$ is 
one half of the square of the geodesic distance between $x$ and $x'$ 
\cite{dewitt}. Therefore, the Green function $G_F$ is 
unique, provided that a
geodesic connecting $x$ and $x'$ is chosen.
When $x$ and $x'$ are sufficiently close to one another,
then there is only one such geodesic. In this
case, the adiabatic expansion \cite{bd,dewitt} of $G_F$ can be used. 
If Riemann normal coordinates are spanned around $x'$, then all 
Green functions can be written as \cite{bd,bunch} 
\begin{equation}\label{green}
G(x,x')=|g(x)|^{-1/4}\int\frac{d^4k}{(2\pi)^4}\, e^{-ik(x-x')}
{\cal G}(k;x') ,
\end{equation} 
where
\begin{equation}\label{adiab}
{\cal G}(k;x')=\sum_{n=0}^{\infty}{\cal G}_n(k;x')
\end{equation}
and ${\cal G}_n(k;x')$ have $n$ derivatives of the metric at $x'$. The
two lowest functions of the adiabatic expansion (\ref{adiab})
are the same as in Minkowski spacetime, 
i.e., ${\cal G}_0=(k^2-m^2)^{-1}$, ${\cal G}_1=0$. 
Depending on the choice of 
the integration contour over $k_0$, from (\ref{green}) one can calculate 
$G_F$, $G^+$, or any other Green function. In particular, $G^+$ 
satisfies
\begin{equation}\label{free_1}
(\nabla^{\mu}\partial_{\mu}+m^2 +\xi R(x))G^+(x,x')=0,
\end{equation}
\begin{equation}\label{free_2}
(\nabla'^{\mu}\partial'_{\mu}+m^2 +\xi R(x'))G^+(x,x')=0 .
\end{equation}
Since $G^+(x,x')$ decreases rapidly with spacelike separation, the 
contributions to (\ref{16}) from large spacelike separations may be
negligible. Therefore, 
for practical calculations, it may be sufficient to calculate 
$G^+(x,x')$ only for small spacelike separations, in which case  
the first few terms of the adiabatic expansion are sufficient.

When $x$ and $x'$ are not close to one another,
then there may be more than one geodesic connecting $x$ and $x'$. 
In this case, $\langle x,s|x',0\rangle$ depends on the choice 
of the geodesic, the point $x$ is not contained in the subspace covered by
the Riemann normal coordinates spanned around $x'$ and the 
adiabatic expansion does not work. Nevertheless, $\langle x,s|x',0\rangle$ 
can be computed for any choice of the 
geodesic connecting $x$ and $x'$. The 
resulting Green function $G^+$ satisfies (\ref{free_1}) and 
(\ref{free_2}).   
To define the exact unique particle current, we need a natural
generalization of $G^+$ to the case with more than one
geodesic connecting $x$ and $x'$. The most natural choice is the
2-point function $\bar{G}^+$ defined as the average over
all geodesics connecting $x$ and $x'$. Assuming that there are $N$
such geodesics, this average is (for any Green function $G$)
\begin{equation}\label{average}
\bar{G}(x,x')=N^{-1} \sum_{a=1}^{N} G(x,x';\sigma_a) ,
\end{equation}
where $G(x,x';\sigma_a)$ is the Green function $G(x,x')$ calculated
with respect to the geodesic $\sigma_a$. 

In the case with a continuous set of geodesics connecting $x$ and $x'$, 
the generalization of (\ref{average}) is more complicated. At the point 
$x'$ we span the flat Minkowski tangent space. Then we perform the 
Wick rotation on the tangent space which transforms it to a Euclidean 
tangent space. Each geodesic emanating from $x'$ is uniquely determined 
by the space angle $\Omega=(\varphi,\vartheta_1,\vartheta_2)$ in the 
Euclidean tangent space because the space angle determines 
the direction of a geodesic emanating from $x'$. Therefore, the measure 
on the set of all geodesics emanating from $x'$ is (see, for example, 
\cite{ryder})
\begin{equation}
\int d\Omega_{x'}=\int_{0}^{2\pi}d\varphi 
\int_{0}^{\pi}{\rm sin}\vartheta_1\, d\vartheta_1
\int_{0}^{\pi}{\rm sin}^2\vartheta_2\, d\vartheta_2,
\end{equation}    
where geodesics emanating in two opposite directions are counted as 
two different geodesics. Of course, not all geodesics emanating from 
$x'$ cross the point $x$, but some of them do. The Green function 
corresponding to a geodesic determined by its space angle $\Omega_{x'}$ 
at $x'$ and crossing the point $x$ is denoted by $G(x,x';\Omega_{x'})$. 
Therefore, we introduce the averaged Green function  
\begin{equation}\label{aver1}
G_{x'}(x,x')=\frac{ \displaystyle\int'd\Omega_{x'} G(x,x';\Omega_{x'}) }
{ \displaystyle\int'd\Omega_{x'}} ,
\end{equation}
where $\int'$ denotes integration only over those space angles for which the
corresponding geodesic crosses the point $x$.
However, a geodesic connecting $x$ and $x'$ can also be specified by its 
angle in the Euclidean tangent space spanned at $x$, resulting
in the averaged Green function
\begin{equation}\label{aver2}                                       
G_{x}(x,x')=\frac{ \displaystyle\int'd\Omega_{x} G(x,x';\Omega_{x}) }
{ \displaystyle\int'd\Omega_{x}} .
\end{equation}
In general, (\ref{aver1}) and (\ref{aver2}) are not equal. To treat 
the points $x$ and $x'$ in a symmetric way, we introduce the 
symmetrized Green function
\begin{equation}
\bar{G}(x,x')=\frac{1}{2}[G_{x}(x,x')+G_{x'}(x,x')] .
\end{equation}
The symmetrized Green function satisfies (\ref{22}), i.e.
\begin{equation}
\bar{G}^*(x,x')=\bar{G}(x',x),
\end{equation}
provided that $G$ is a Green function such that 
$G_{x}^*(x,x')=G_{x}(x',x)$ and $G_{x'}^*(x,x')=G_{x'}(x',x)$. 

Now the most natural choice for the particle current in arbitrary 
curved background is 
\begin{equation}\label{current_gen}
j_{\mu}(x)=\cint'^{\nu} \frac{1}{2} \{\bar{G}^+(x,x')\apm \; \apn
\phi(x)\phi(x') + {\rm h.c.} \}.
\end{equation}
Since the function $\bar{G}^+(x,x')$ 
satisfies (\ref{free_1}) and (\ref{free_2}), it
follows that the particle current (\ref{current_gen}) 
does not depend on the choice of $\Sigma$ 
and is conserved 
\begin{equation}
\nabla^{\mu}j_{\mu}(x)=0,
\end{equation}
provided that the field satisfies (\ref{8}). 
Note that the corresponding definition of
particles does not
always correspond to the quantities detected
by ``particle detectors" of the Unruh-DeWitt type \cite{unruh,dewitt2}.
Instead, the number of particles resulting from (\ref{current_gen})
is determined by a well-defined
hermitian operator that does not require a model of a particle
detector, just as is the case for all other observables
in quantum mechanics. Moreover, since the observable $\phi(x)$
in (\ref{current_gen}) does not require
a choice of representation of field algebra, the definition
of particles based on $\bar{G}^+$ does not require the choice of
representation either. This allows us to treat the particles in
the framework of algebraic quantum field theory in curved
background \cite{wald}.
Furthermore, as noted by Unruh
\cite{unruh}, only one definition of particles can correspond to the
real world, in the sense that their stress-energy contributes to
the gravitational field. Since the definition of particles based on
$\bar{G}^+$ is universal, unique, and really covariant, it might
be that these are the particles that correspond to the real world. 
Similarly, from the discussion in Sec.~\ref{SEC2},
it seems that these particles might correspond
to the objects detected by real detectors 
(such as a Wilson chamber or
a Geiger-M\"{u}ller counter) in real experiments. 
If such an interpretation of these particles is correct, then
classical gravitational backgrounds do not produce real particles, 
in agreement with some other results 
\cite{padmprl,belin,nikol1,nikol2,nikol3}. 
Note that this does not necessarily imply that 
black holes do not radiate. It is possible that {\em quantum} gravity 
provides a correct mechanism for black hole radiation. 
Whatever the mechanism of the particle creation near the 
horizon might be, one should expect an approximately 
thermal distribution of escaped particles as seen by a distant observer. 
The thermal distribution is a
consequence of the exponential red shift and can be understood
even by classical physics \cite{pad3,pad4}, without any assumption
on the physical mechanism that causes particle creation near the
horizon.

\subsection{Covariance of the time evolution}\label{COVAR}

As we have seen,
when particle production occurs, then the function $W(x,x')$ or the 
field $\phi(x')$ does not satisfy the ``free" equation of motion 
(\ref{21'}) or (\ref{8}), respectively. Consequently, the current 
$j_{\mu}(x)$ depends not only on the point $x$, but also on the 
choice of $\Sigma$ on which $x$ lies. This suggests that
there exists a preferred foliation of spacetime. 
Such a result should not be
surprising because many other aspects of quantum theory
seem to require the existence of a preferred time
coordinate. However, it may be surprising that,
in our analysis, the need for 
a preferred time coordinate has emerged from the requirement of 
general covariance (of the concept od particles). Below we 
argue that the principle of general covariance (together with some 
other natural principles) may help us to determine this 
preferred time coordinate.
 
The time evolution law (\ref{dUdt}) written in the Heisenberg picture is
\begin{equation}\label{dudtheis}
i\frac{d}{dt}U(t)=\tilde{H}(t)U(t) ,
\end{equation}
where (see (\ref{Hint}))
\begin{equation}\label{hamilt}
\tilde{H}(t)=\int_{\Sigma(t)} d^3x \, |g|^{1/2} {\cal H}(x),
\end{equation}
${\cal H}=T^0_0$, and
\begin{equation}
T^{\mu}_{\nu}=\frac{\partial {\cal L}}{\partial(\partial_{\mu}\phi)}
\partial_{\nu}\phi-g^{\mu}_{\nu}{\cal L}.
\end{equation}
Similarly, in the Schr\"{o}dinger picture
\begin{equation}            
i\frac{d}{dt}\Psi[\phi,t)=\tilde{H}\Psi[\phi,t) ,   
\end{equation}
where $\Psi[\phi,t)$ is a functional with respect to $\phi({\bf x})$ and 
a function with respect to $t$, while the Hamiltonian $\tilde{H}$ is  
given by (\ref{hamilt}) in which $\tilde{\pi}({\bf x})=
-i\delta/\delta\phi({\bf x})$ (see (\ref{canon_pi})). 
This time-evolution law is not manifestly covariant. This is because $t$ and 
$\tilde{H}$ are not scalars. This law would become covariant if $dt\,\tilde{H}$ 
could be replaced with $d\tau\,H$, where
\begin{equation}\label{eq103}
d\tau=n_{\mu}dx^{\mu}=\frac{g_{0\mu}}{\sqrt{g_{00}}}dx^{\mu},
\end{equation}
\begin{equation}\label{eq104}
H=\cint^{\mu}n^{\nu}T_{\mu\nu}=\int_{\Sigma}d^3x\,|g^{(3)}|^{1/2}
\,\frac{g_{0\alpha}T_0^{\alpha}}{g_{00}}
\end{equation}
are scalars. It is easy to see that $dt\,\tilde{H}=d\tau\,H$ 
for any $T^{\mu}_{\nu}$ if and only if 
\begin{equation}
g_{0i}=0 \;\;\; {\rm and} \;\;\; \partial_ig_{00}=0.
\end{equation}
In this case, $d\tau=\sqrt{g_{00}(t)}dt$, so the metric takes the form
\begin{equation}\label{gauss}
ds^2=d\tau^2-g_{ij}(\tau,{\bf x})dx^idx^j .
\end{equation}
The metric can always be written in this form. The coordinates 
for which the metric takes the form (\ref{gauss}) are known as 
Gaussian coordinates \cite{wein}.

The results above can be reinterpreted in the following way. 
The covariant time evolution determined by (\ref{eq103}) and 
(\ref{eq104}) depends on the foliation of spacetime into Cauchy 
spacelike hypersurfaces. The choice of foliation corresponds to the choice 
of coordinates for which $t={\rm constant}$ on $\Sigma$. Therefore, we 
can determine the time evolution uniquelly by giving a preferred status to a
set of coordinates. The Gaussian coordinates are special because only in 
these coordinates the time-evolution law takes the canonical form 
(\ref{dudtheis}), required by the usual rules of quantum mechanics.
This shows that the requirement of covariance of the 
time-evolution law leads to Gaussian coordinates as preferred coordinates.

However, for a given spacetime, the Gaussian coordinates are not unique. 
One needs an additional criterion for choosing them. 
For example, a reasonable 
additional criterion is the requirement that there should be no coordinate 
singularity in these coordinates, but this requirement still 
does not make the Gaussian coordinates unique. 
An interesting attempt to choose the Gaussian coordinates in a 
unique way is given in \cite{capri1,capri2}. However, in the approach
of \cite{capri1,capri2}, 
the coordinate singularities such as horizons are not avoided. 
In the case of a Schwarzschild
black hole, a particularly interesting choice of the 
Gaussian coordinates is the Lemaitre coordinates, in which there is no 
coordinate singularity. In these coordinates, the Schwarzschild black 
hole is time dependent, the Hawking radiation looks thermal 
to a distant outside observer, but the absence of the horizon implies 
a unitary time evolution which resolves the 
black-hole information paradox \cite{melnik}.

\section{Complex scalar field}\label{SEC4}

To explore the similarity between the current of particles and that
of charge, we generalize some results for hermitian scalar fields to the case
of complex scalar fields. We do 
not discuss those aspects the generalization of which is obvious. 
We also study the scalar QED, i.e., the interaction 
of complex scalar fields with classical 
and quantum electromagnetic fields. 

\subsection{Particle and charge currents}\label{C1}

The complex field $\phi$ 
can be written in terms of hermitian fields $\phi_1$ and
$\phi_2$ as 
\begin{equation}
\phi=\frac{\phi_1 +i\phi_2}{\sqrt{2}}.
\end{equation}
The independent fields $\phi_1$ and $\phi_2$ satisfy the same 
canonical commutation relations 
as the hermitian scalar field of Sec.~\ref{HERMIT}, 
with an additional property $[\phi_1(x),\phi_2(x')]=0$. Therefore, the 
canonical commutation relations can be written as
\begin{equation}\label{comm1}
[\phi^{\dagger}(x),\phi(x')]_{\Sigma}=
[\partial_0\phi^{\dagger}(x),\partial'_0\phi(x')]_{\Sigma}=0,
\end{equation}
\begin{equation}\label{comm2}
\tilde{n}^0(x') [\phi^{\dagger}(x),\partial'_0\phi(x')]_{\Sigma}
=i \delta^3({\bf x}-{\bf x}') .
\end{equation}
The complex field $\phi$ satisfies (\ref{8}) 
and can be expanded as
\begin{eqnarray}\label{25}
\phi(x)=\sum_{k}a_k f_k(x)+ b^{\dagger}_k f^*_k(x)  ,
\nonumber \\
\phi^{\dagger}(x)=\sum_{k}a^{\dagger}_k f^*_k(x)+ b_k f_k(x)  .
\end{eqnarray}
The operators $a_k$ ($a^{\dagger}_k$) destroy (create) particles, while 
$b_k$ ($b^{\dagger}_k$) destroy (create) antiparticles. They satisfy 
the corresponding algebra of lowering and raising operators and can be 
expressed as
\begin{eqnarray}
a_k=(f_k,\phi), \;\;\; b^{\dagger}_k=-(f^*_k,\phi), \nonumber \\
a^{\dagger}_k=-(f^*_k,\phi^{\dagger}), \;\;\; b_k=(f_k,\phi^{\dagger}).
\end{eqnarray}
We introduce two global quantities:
\begin{equation}\label{26}
N^{(\pm)}=\sum_{k}a^{\dagger}_k a_k \pm b^{\dagger}_k b_k .
\end{equation}
Here $N^{(+)}$ is the total number of particles, while $N^{(-)}$ is
the total charge. In a way similar to the case of 
hermitian field, we find the covariant expression
\begin{equation}\label{27}
N^{(\pm)}=\cint^{\mu} j^{(\pm)}_{\mu}(x) ,
\end{equation}
where
\begin{equation}\label{28}
j^{(\pm)}_{\mu}(x)  =  \cint'^{\nu} \frac{1}{2} \{W(x,x')\apm \; \apn
[\phi^{\dagger}(x)\phi(x') \pm \phi(x)\phi^{\dagger}(x')] +
{\rm h.c.} \} 
\end{equation}
and $W(x,x')$ is given by (\ref{18r}).
Obviously, $j^{(+)}_{\mu}$ should be interpreted as a current of 
particles, while $j^{(-)}_{\mu}$ should be interpreted as a current of 
charge. These two currents can also be written as 
\begin{equation}
j^{(\pm)}_{\mu}=j^{(P)}_{\mu} \pm j^{(A)}_{\mu},
\end{equation}
where
\begin{equation}\label{curP}
j^{(P)}_{\mu}(x)=\cint'^{\nu} \frac{1}{2} \{W(x,x')\apm \; \apn
\phi^{\dagger}(x)\phi(x') +{\rm h.c.} \} ,
\end{equation}
\begin{equation}\label{curA}
j^{(A)}_{\mu}(x)=\cint'^{\nu} \frac{1}{2} \{W(x,x')\apm \; \apn
\phi(x)\phi^{\dagger}(x') +{\rm h.c.} \}
\end{equation}
are the currents of particles and antiparticles, respectively.  
It is also instructive to write the currents in terms of hermitian 
fields $\phi_1$ and $\phi_2$.
The current of particles is a sum of two currents of the form (\ref{16});
\begin{equation}
j^{(+)}_{\mu}(x)  =  \cint'^{\nu} \frac{1}{2} \{W(x,x')\apm \; \apn
[\phi_1(x)\phi_1(x') + \phi_2(x)\phi_2(x')] +
{\rm h.c.} \} ,
\end{equation}
while $j^{(-)}_{\mu}$ takes a form from which it is manifest 
that the current of charge does not exist in the case of only one 
hermitian field;
\begin{equation}
j^{(-)}_{\mu}(x)  =  \cint'^{\nu} \frac{1}{2} \{W(x,x')\apm \; \apn
i[\phi_1(x)\phi_2(x') - \phi_2(x)\phi_1(x')] +
{\rm h.c.} \} .
\end{equation}

There is one important difference between the currents $j^{(+)}_{\mu}$ and 
$j^{(-)}_{\mu}$. Similarly to the particle current (\ref{16}), the 
particle current $j^{(+)}_{\mu}(x)$ possesses the non-local features 
related to the integration over $\Sigma$ on which $x$ lies. On the 
other hand,  
the apparent non-local features of $j^{(-)}_{\mu}(x)$ really do not
exist, because, by using the canonical commutation relations
(\ref{comm1})-(\ref{comm2}) and 
Eqs.~(\ref{19})-(\ref{20}), 
the integration over $d\Sigma'^{\nu}$ can be performed. 
This cannot be done for $j^{(+)}_{\mu}$ because, owing to the 
different sign, certain terms that cancel in $j^{(-)}_{\mu}$ 
do not cancel in $j^{(+)}_{\mu}$.
Applying the commutation relations (\ref{comm1})-(\ref{comm2}) to 
$j^{(-)}_{\mu}$ given by (\ref{28}) such that $\phi^{\dagger}$ always  
comes to the left and $\phi$ to the right, a 
straightforward calculation that exploits Eqs.~(\ref{19})-(\ref{20}) yields
\begin{equation}\label{30.1}
j^{(-)}_{\mu}(x)  =  i\phi^{\dagger}(x)\apm \phi(x) 
  + \cint'^{\nu} W(x,x')\apm \; \apn W(x',x) ,
\end{equation}
so all non-local features are contained in the second term that
does not depend on $\phi$. 
Similarly, applying the commutation relations such that 
$\phi^{\dagger}$ always 
comes to the right and $\phi$ to the left, 
$j^{(-)}_{\mu}$ can be written as 
\begin{equation}\label{30.2}
j^{(-)}_{\mu}(x)  =  -i\phi(x)\apm \phi^{\dagger}(x) 
  - \cint'^{\nu} W(x',x)\apm \; \apn W(x,x') .
\end{equation}
From the antisymmetry of the operators $\apm$ and $\apn\,$, it follows that 
the integrals appearing in (\ref{30.1}) and (\ref{30.2}) are equal.
Therefore, by summing 
(\ref{30.1}) and (\ref{30.2}) and dividing the sum by 2, we find a purely local
expression
\begin{equation}\label{charge_current}
j^{(-)}_{\mu}(x)=\frac{i}{2}[\phi^{\dagger}(x)\apm \phi(x)-
\phi(x)\apm \phi^{\dagger}(x)] .
\end{equation}
This is the usual form of the charge current, often derived as a Noether 
current resulting from the global U(1) invariance of the Lagrangian.
 
Using (\ref{25}), one can show that (\ref{charge_current}) is normally 
ordered, i.e., that 
\begin{equation}\label{normal_ord}
\langle 0|j^{(-)}_{\mu}| 0\rangle =0.
\end{equation}
Alternatively, one can prove (\ref{normal_ord}) 
by using (\ref{25}), (\ref{18r}), and (\ref{11}) to show that 
\begin{equation}
\langle 0|i\phi^{\dagger}(x)\!\apm\! \phi(x)| 0\rangle
=-\cint'^{\nu} W(x,x')\apm \; \apn W(x',x),
\end{equation}
which implies that (\ref{30.1}) can be written as
\begin{equation}
j^{(-)}_{\mu} = i\phi^{\dagger}\apm \phi
-\langle 0|i\phi^{\dagger}\apm \phi| 0\rangle .
\end{equation} 

Using (\ref{18r}), (\ref{25}), (\ref{11}), and $[a_k,b_{k'}]=0$, 
the currents (\ref{28}), (\ref{curP}), and (\ref{curA}) 
can be written in a form similar to (\ref{22_mu}). We find
\begin{equation}\label{c1}
j_{\mu}^{(P)}=i\sum_{k,k'}f_k^* \apm f_{k'} a_k^{\dagger}a_{k'}
+j_{\mu}^{{\rm mix}},
\end{equation}
\begin{equation}\label{c2}
j_{\mu}^{(A)}=i\sum_{k,k'}f_k^* \apm f_{k'} b_k^{\dagger}b_{k'}
-j_{\mu}^{{\rm mix}},
\end{equation}
\begin{equation}\label{c3}
j_{\mu}^{(+)}=i\sum_{k,k'}f_k^* \apm f_{k'} (a_k^{\dagger}a_{k'}
+b_k^{\dagger}b_{k'}),
\end{equation}
\begin{equation}\label{c4}
j_{\mu}^{(-)}=i\sum_{k,k'}f_k^* \apm f_{k'} (a_k^{\dagger}a_{k'}
-b_k^{\dagger}b_{k'}) +2j_{\mu}^{{\rm mix}},
\end{equation}
where the mixed hermitian current $j_{\mu}^{{\rm mix}}$ 
mixes (in products) the particle operators $a_k$, $a_k^{\dagger}$ 
with the antiparticle operators $b_k$, $b_k^{\dagger}$:
\begin{equation}
j_{\mu}^{{\rm mix}}=\frac{i}{2}\sum_{k,k'}
(f_k \apm f_{k'} a_{k'}b_k + f_k^* \apm f_{k'}^*
a_{k}^{\dagger}b_{k'}^{\dagger}) .
\end{equation}  
Now we see that
\begin{equation}
j^{(+)}_{\mu}| 0\rangle =0,
\end{equation}
while the currents (\ref{c1}), (\ref{c2}), and (\ref{c4}) have not this 
property. In the case of Minkowski spacetime, Eq.~(\ref{23}) generalizes 
to
\begin{equation}\label{34}
\tilde{J}^{(\pm)}_{\mu}\equiv \int d^3x \, j^{(\pm)}_{\mu}(x)=\sum_{{\bf k}}
\frac{k_{\mu}}{\omega_{{\bf k}}} (a^{\dagger}_{{\bf k}}
a_{{\bf k}} \pm b^{\dagger}_{{\bf k}} b_{{\bf k}}).
\end{equation}

From (\ref{8}) and (\ref{21}) it follows that 
the currents (\ref{28}) are conserved:
\begin{equation}
\nabla^{\mu}j^{(\pm)}_{\mu}=0.
\end{equation}
When a non-gravitational interaction is described by (\ref{35}), then 
$j^{(+)}_{\mu}$ satisfies an equation similar to (\ref{36}), i.e.
\begin{equation}\label{complex_J_prod}
\nabla^{\mu}j^{(+)}_{\mu}(x) = \cint'^{\nu} \frac{1}{2}\{
W(x,x')\apn [J^{\dagger}(x)\phi(x') + J(x)\phi^{\dagger}(x')]
 +{\rm h.c.}\} .
\end{equation}
On the other hand,
$j^{(-)}_{\mu}$ is conserved again,
provided that the Lagrangian possesses a global 
U(1) symmetry. The conservation of $j^{(-)}_{\mu}$ 
is related to the fact that the second term in (\ref{30.1}) does 
not depend on the interaction described by $J$.  

When a natural choice of the modes $f_k$ does not exist, then one can 
redefine the particle current $j^{(+)}_{\mu}$ (\ref{28}) by replacing 
(\ref{18r}) with  
(\ref{Wcreate}) or $\bar{G}^{+}(x,x')$ from Sec.~\ref{NATURAL}.  

\subsection{The generalization to the electromagnetic background}

Let us generalize the definition of currents $j^{(\pm)}_{\mu}$
to the case in which a classical 
electromagnetic background $\bar{A}_{\mu}(x)$ is also present.
Under a gauge transformation
\begin{equation}
\bar{A}'_{\mu}(x)=\bar{A}_{\mu}(x)+e^{-1}\partial_{\mu}\lambda(x),
\end{equation}
the scalar field transforms as
\begin{equation}\label{phi_gauge}
\phi'(x)=\phi(x) e^{-i\lambda(x)} , \;\;\; 
\phi'^{\dagger}(x)=\phi^{\dagger}(x) e^{i\lambda(x)}.
\end{equation}
The covariant derivatives 
\begin{equation}\label{covar_deriv}
D_{\mu}=\nabla_{\mu}+ie\bar{A}_{\mu}, \;\;\;
D^*_{\mu}=\nabla_{\mu}-ie\bar{A}_{\mu}
\end{equation}
are covariant with respect to both the gauge 
transformations and the general-coordinate 
transformations. The quantities $D_{\mu}\phi$ and $D^*_{\mu}\phi^*$ 
transform as $\phi$ and $\phi^*$ in (\ref{phi_gauge}), respectively.
The fields $\phi$ and $\phi^*$ satisfy the equations of motion 
\begin{eqnarray}\label{g_eq_m}
& (D^{\mu}D_{\mu}+m^2 +\xi R)\phi=0, & \nonumber \\
& (D^{*\mu}D^*_{\mu}+m^2 +\xi R)\phi^{\dagger}=0. &
\end{eqnarray}
The solution can be expanded as
\begin{eqnarray}\label{gexp}
& \phi(x)=\displaystyle\sum_{k}a_k f_k(x)
+ b^{\dagger}_k g^*_k(x), & \nonumber \\
& \phi^{\dagger}(x)=\displaystyle\sum_{k}a_k^{\dagger} f^*_k(x)
+ b_k g_k(x). &
\end{eqnarray}
When $\bar{A}_{\mu}=0$, then $f_k=g_k$. In general, $f_k$ and $g^*_k$ 
transform as $\phi$ under a gauge transformation, while 
$f^*_k$ and $g_k$ transform as $\phi^{\dagger}$.  
Introducing the ``antisymmetric" covariant derivative $\gapm$ defined by
\begin{equation}
a\gapm b=aD_{\mu}b-(D^*_{\mu}a)b,
\end{equation}
we define two different scalar products:
\begin{eqnarray}
& (a,b)=i\displaystyle\cint^{\mu} a^* \gapm b , & \nonumber \\
& (a,b)_*=i\displaystyle\cint^{\mu} a^* \gapms b . &
\end{eqnarray}
These scalar products are gauge invariant, provided that $a$ and $b$
transform in the same way under the gauge transformation, i.e., either 
as $\phi$ or as $\phi^{\dagger}$. When $a$ commutes with $b$, then
\begin{equation}
a\gapms b=-b\gapm a.
\end{equation}
Consequently, in the commuting case, the two scalar products are related as
\begin{equation}
(a,b)_*=-(b^*,a^*) .
\end{equation}
The modes $f_k$ and $g_k$ satisfy
\begin{eqnarray}
& (f_k,f_{k'})=-(f^*_k,f^*_{k'})_*=\delta_{kk'}, & \nonumber \\
& (g_k,g_{k'})_*=-(g^*_k,g^*_{k'})=\delta_{kk'}, & \nonumber \\
& (f_k,g^*_{k'})=(g_k,f^*_{k'})_*=0 . &
\end{eqnarray}
Therefore, the operators $a_k$ and $b_k$ satisfy the usual 
algebra of lowering and raising operators and can be
written as
\begin{eqnarray}
& a_k=(f_k,\phi), \;\;\; b^{\dagger}_k=-(g^*_k,\phi), & \nonumber \\
& a^{\dagger}_k=-(f^*_k,\phi^{\dagger})_*, \;\;\; 
b_k=(g_k,\phi^{\dagger})_*. &
\end{eqnarray}
The vacuum $|0\rangle$ is defined by
\begin{equation}\label{g_vacuum}
a_k |0\rangle = b_k |0\rangle =0.
\end{equation}
 
We introduce two 2-point functions:
\begin{eqnarray}\label{gW12}
& W^{(P)}(x,x')=\displaystyle\sum_k f_k(x)f^*_k(x'), & \nonumber \\
& W^{(A)}(x,x')=\displaystyle\sum_k g_k(x)g^*_k(x'). &
\end{eqnarray}
From (\ref{gexp}) and the fact that the operators $a_k$ and $a^{\dagger}_k$ 
are related to particles while $b_k$ and $b^{\dagger}_k$ are related to 
antiparticles, we see that $W^{(P)}$ is related to the propagator of 
particles, while $W^{(A)}$ is related to the propagator of antiparticles.
In general, $W^{(P)}\neq W^{(A)} $, which corresponds to the fact that particles 
and antiparticles propagate in a different way, provided that a background 
electromagnetic field exists. When $\bar{A}_{\mu}=0$, then 
$W^{(P)}=W^{(A)}=W$, where $W$ is the 2-point function used in Sec.~\ref{C1}. 
This means that particles
and antiparticles propagate in the same way when a background
electromagnetic field does not exist. When 
$\bar{F}_{\mu\nu}\equiv\partial_{\mu}\bar{A}_{\nu}-\partial_{\nu}\bar{A}_{\mu}
=0$ but $\bar{A}_{\mu}=e^{-1}\partial_{\mu}\lambda\neq 0$, then
\begin{eqnarray}
& W^{(P)}(x,x')=e^{-i[\lambda(x)-\lambda(x')]}W(x,x'), & \nonumber \\
& W^{(A)}(x,x')=e^{i[\lambda(x)-\lambda(x')]}W(x,x'), &
\end{eqnarray}
so the two 2-point functions differ only by a phase, which should not have 
any physical effects.
The functions $W^{(P)}$ and $W^{(A)}$ satisfy
\begin{eqnarray}\label{eq_m_W}
& (D^{\mu}D_{\mu}+m^2 +\xi R(x))W^{(P)}(x,x')=0, & \nonumber \\
& (D'^{*\mu}D'^*_{\mu}+m^2 +\xi R(x'))W^{(P)}(x,x')=0, & \nonumber \\
& (D^{*\mu}D^*_{\mu}+m^2 +\xi R(x))W^{(A)}(x,x')=0, & \nonumber \\
& (D'^{\mu}D'_{\mu}+m^2 +\xi R(x'))W^{(A)}(x,x')=0. &
\end{eqnarray}
Therefore, $W^{(P)}$ and $W^{(A)}$ transform to each other when $\bar{A}_{\mu}$ 
transforms to $-\bar{A}_{\mu}$, which corresponds to the transformation 
$\bar{F}_{\mu\nu} \rightarrow -\bar{F}_{\mu\nu}$.
From (\ref{gexp}) it follows that $W^{(P)}$ and $W^{(A)}$ are equal to the 
Wightman functions
\begin{eqnarray}
& W^{(P)}(x,x')=\langle 0| \phi(x)\phi^{\dagger}(x') |0\rangle, & \nonumber \\
& W^{(A)}(x,x')=\langle 0| \phi^{\dagger}(x)\phi(x') |0\rangle, &
\end{eqnarray}
while from  (\ref{gW12}) we find
\begin{equation}
W^{(P)*}(x,x')=W^{(P)}(x',x), \;\;\; W^{(A)*}(x,x')=W^{(A)}(x',x),
\end{equation}
Under gauge transformations, $W^{(P)}(x,x')$ transforms as
$\phi(x)\phi^{\dagger}(x')$, while $W^{(A)}(x,x')$ transforms as
$\phi^{\dagger}(x)\phi(x')$.
From the canonical commutation relations 
\begin{equation}\label{gcomm1}
[\phi^{\dagger}(x),\phi(x')]_{\Sigma}=
[D^*_0\phi^{\dagger}(x),D'_0\phi(x')]_{\Sigma}=0,
\end{equation}
\begin{equation}\label{gcomm2}
\tilde{n}^0(x') [\phi^{\dagger}(x),D'_0\phi(x')]_{\Sigma}
=i \delta^3({\bf x}-{\bf x}') ,
\end{equation}
we find that $f_k$ and $g_k$ 
are functions such that
\begin{eqnarray}\label{g789}
& [W^{(P)}(x,x')-W^{(A)}(x',x)]_{\Sigma}=0, & \nonumber \\
& D_0D'^*_0[W^{(P)}(x,x')-W^{(A)}(x',x)]_{\Sigma}=0, & \nonumber \\
& \tilde{n}^0(x') D'^*_0[W^{(P)}(x,x')-W^{(A)}(x',x)]_{\Sigma}=
i\delta^3({\bf x}-{\bf x}'). &
\end{eqnarray}

Now we introduce the global quantities:
\begin{equation}
N^{(\pm)}=\sum_{k}a^{\dagger}_k a_k \pm b^{\dagger}_k b_k  .
\end{equation}
Writing this in the covariant form (\ref{27}), we find
\begin{eqnarray}\label{g28}
j^{(\pm)}_{\mu}(x) & = & \cint'^{\nu} \frac{1}{2} \{ [W^{(P)}(x,x')\gapms \; 
\gapn \phi^{\dagger}(x)\phi(x')
\nonumber \\
 & & \pm W^{(A)}(x,x')\gapm \; \gapns \phi(x)\phi^{\dagger}(x')] +{\rm h.c.} \}.
\end{eqnarray}
Although $W^{(A)}$, $W^{(P)}$, $\phi$, and $\phi^{\dagger}$ depend on gauge, the
currents (\ref{g28}) do not depend on gauge. Therefore, our definition 
of particles and that of charge are gauge invariant. Eq.~(\ref{g28}) is the 
gauge-invariant generalization of (\ref{28}). 

In a way similar to the case $\bar{A}_{\mu}=0$, it can be shown that
$j^{(-)}_{\mu}$ can be written in a purely local form 
\begin{eqnarray}\label{g_cur_local}
j^{(-)}_{\mu} & = & i\phi^{\dagger}\gapm\phi -
 \langle 0| i\phi^{\dagger}\gapm\phi |0\rangle \nonumber \\
& = & -i\phi\gapms\phi^{\dagger} +
 \langle 0| i\phi\gapms\phi^{\dagger} |0\rangle .
\end{eqnarray}
This is the usual form of the charge current, often derived 
from the gauge invariance of the Lagrangian. 

From (\ref{eq_m_W}) and 
(\ref{g_eq_m}) it follows that both currents are conserved:
\begin{equation}
\nabla^{\mu}j^{(\pm)}_{\mu}=0.
\end{equation}
Of course, the conservation of the charge current 
$j^{(-)}_{\mu}$ was expected because 
it is well known that (\ref{g_cur_local}) is conserved. However, 
the conservation of $j^{(+)}_{\mu}$ is less trivial because it 
implies that 
the background electromagnetic field described by 
$\bar{A}_{\mu}$ does not produce particles
(particle-antiparticle pairs), provided that a unique 
(or a preferred) vacuum defined by (\ref{g_vacuum}) exists. 

In general, a natural definition of the vacuum 
(\ref{g_vacuum}) does not exist because the natural choice of the modes 
$f_k$ and $g_k$ does not exist. In this case, one can introduce different 
modes that are natural at different times, similarly as in 
Sec.~\ref{PPGB}. In a time-dependent background $\bar{A}_{\mu}$, 
the natural modes at a given time can be 
chosen such that they are positive-frequency modes at that time 
\cite{padmprl}. Alternatively, in a time-independent but 
space-dependent background $\bar{A}_{\mu}$, one can introduce  
``in" and ``out" modes, the naturalness of which is related to 
the sign of the wave vector ${\bf k}$ in asymptotic regions \cite{man}. 
In this way, one can describe the pair creation by the electromagnetic 
background in a local way, similarly as in Sec.~\ref{PPGB}. As in  
the case of particle creation by the gravitational background 
studied in Sec.~\ref{PPGB}, it is easy to see that 
this local description of pair creation is equivalent to the usual 
global description based on the Bogoliubov transformation.

However, the sign of the frequency and that of the wave vector are not  
gauge-invariant concepts. In some cases, such as 
$\bar{F}_{\mu\nu}=0$ or a time-independent 
background electric field, some gauges may seem ``natural", but a 
general recipe for choosing the gauge does not exist. Therefore, 
by using the Bogoliubov-transformation method,  
it is not possible, even in principle, to calculate the distribution 
of produced pairs in an {\em arbitrary} electromagnetic background. 
 
The following remark also illustrates why the Bogoliubov-transformation 
method is problematic.
For $\bar{F}_{\mu\nu}=0$, the
``natural" gauge is $\bar{A}_{\mu}=0$, which defines the
``natural" vacuum. However, by using a different gauge, one can
obtain that this vacuum is a many-particle state. In particular,
an appropriate gauge leads to a thermal distribution of
particles in the ``natural" vacuum, leading to an
electromagnetic analog of the Rindler quantization 
and the Unruh effect. On the other hand,
physical quantities should not depend on gauge, which again 
raises doubts about the appropriateness of the 
Bogoliubov-transformation method. 

Now we see that just as the particle current based on (\ref{Wcreate}) 
is not really general-covariant, a similar modification of 
(\ref{g28}) is not really gauge-invariant.
To define particles in a unique way,  
one needs a universal natural 
choice of the functions $W^{(P)}$ and $W^{(A)}$ using a method that does not 
require a particular choice of the modes $f_k$ and $g_k$. Again, such a 
choice exists and is a generalization of the choice in 
Sec.~\ref{NATURAL}. By introducing the Feynman Green functions 
$G^{(P)}_F$ and $G^{(A)}_F$ calculated by using Schwinger's method 
\cite{schw,step,paren,tsam}, we propose
\begin{equation}\label{W=G}
W^{(P)}(x,x')=G^{(P)+}(x,x'), \;\;\; W^{(A)}(x,x')=G^{(A)+}(x,x').
\end{equation}
(Note that the Green functions are calculated for Minkowski 
spacetime in the references cited above, but the formalism can be 
generalized to curved spacetime as well.) 
Here $G^{(P)+}$ and $G^{(A)+}$ satisfy (\ref{eq_m_W}), i.e.
\begin{eqnarray}\label{eq_m_G}
& (D^{\mu}D_{\mu}+m^2 +\xi R(x))G^{(P)+}(x,x')=0, & \nonumber \\
& (D'^{*\mu}D'^*_{\mu}+m^2 +\xi R(x'))G^{(P)+}(x,x')=0, & \nonumber \\
& (D^{*\mu}D^*_{\mu}+m^2 +\xi R(x))G^{(A)+}(x,x')=0, & \nonumber \\
& (D'^{\mu}D'_{\mu}+m^2 +\xi R(x'))G^{(A)+}(x,x')=0, &
\end{eqnarray}
and are calculated from the corresponding Feynman Green functions 
that satisfy similar equations, but with the right-hand sides 
proportional to $\delta^4(x-x')$. Note that in the literature 
\cite{schw,step,paren,tsam}, usually only one Feynman Green function  
corresponding to $G^{(P)}_F$ is considered and is called simply 
$G_F$. However, there are actually two Feynman Green functions 
$G^{(P)}_F$ and $G^{(A)}_F$ that satisfy similar equations, but with 
$\bar{A}_{\mu}$ replaced with $-\bar{A}_{\mu}$.      
      
From (\ref{W=G}), (\ref{eq_m_G}), and (\ref{g_eq_m}) 
it follows that the currents (\ref{g28}) are conserved. 
In particular, this implies that the most natural definition 
of the particle current in an electromagnetic background leads 
to the result that classical electromagnetic backgrounds 
do not produce particles. This is in agreement with some 
other results \cite{nikol1,nikol2,nikol3}. 
It is interesting to note that a different gauge-invariant formalism 
based on the same Green function $G_F$ as above leads to a result 
that a classical background static electric field produces 
particle-antiparticle pairs \cite{schw}. 
However, when
the formalism of \cite{schw} is naturally modified so that the     
integration contour over a scalar variable
remains on the real axis (see (\ref{integral_s})),
then this formalism does not lead to pair production.
As argued in \cite{nikol3}, such a modified formalism might be 
the correct one.

\subsection{The generalization to the quantum electromagnetic interaction}

As we have seen, the result that the interaction with a classical 
electromagnetic background does not lead to a change 
of the number of particles  
is closely related to gauge invariance. However, it is known 
from many experiments that the electromagnetic interaction may cause 
a change of the number of charged particles. The best known example 
is the annihilation of a particle-antiparticle pair to a pair of 
photons. Such processes are successfully described by Feynman diagrams 
that result from a theory in which the electromagnetic field 
is also quantized. Physical arguments explaining why only a quantum 
electromagnetic interaction can consistently describe a change 
of the number of particles are given in \cite{nikol3}. Here we 
give a formal explanation of this.

Let $\hat{A}_{\mu}$ be the total operator of the electromagnetic 
field. It can be written as 
\begin{equation}
\hat{A}_{\mu}=A_{\mu}+\bar{A}_{\mu},
\end{equation}
where $A_{\mu}$ is the operator such that 
\begin{equation}
\langle 0|A_{\mu}|0 \rangle =0,
\end{equation}
from which it follows that
\begin{equation}
\langle 0|\hat{A}_{\mu}|0 \rangle =\bar{A}_{\mu}.
\end{equation}
When $A_{\mu}$ does not interact with other fields, then it can 
be expanded in the usual way in terms of creation and annihilation 
operators. A gauge transformation 
\begin{equation}
\hat{A}'_{\mu}=\hat{A}_{\mu}+e^{-1}\partial_{\mu}\lambda,
\end{equation}
where $\lambda(x)$ is a c-number function, can be written as
\begin{eqnarray}\label{q_g_transf}
& A'_{\mu}=A_{\mu}, & \nonumber \\
& \bar{A}'_{\mu}=\bar{A}_{\mu}+e^{-1}\partial_{\mu}\lambda. &
\end{eqnarray}
Therefore, the operator $A_{\mu}$ is invariant with respect to 
the classical gauge transformations (\ref{q_g_transf}). The generalization 
of (\ref{covar_deriv}) is  
\begin{equation}\label{q_covar_deriv}
\hat{D}_{\mu}=\nabla_{\mu}+ie\hat{A}_{\mu}=
D_{\mu}+ieA_{\mu}.
\end{equation}
The currents $j^{(\pm)}_{\mu}$ take the same form 
(\ref{g28}) with the same covariant derivative $\gapm$ and 
with the same functions $W^{(P)}$ and $W^{(A)}$ satisfying 
Eqs.~(\ref{eq_m_W}). However, in these currents, the field $\phi$ satisfies 
\begin{equation}\label{q_g_eq_m}
(\hat{D}^{\mu}\hat{D}_{\mu}+m^2 +\xi R)\phi=0, 
\end{equation}
and similarly for $\phi^{\dagger}$. Eq.~(\ref{q_g_eq_m}) can also be written 
in a form similar to (\ref{35}):
\begin{equation}\label{complex_J}
(D^{\mu}D_{\mu} +m^2 +\xi R)\phi=J,
\end{equation}
where
\begin{equation}
J=[e^2 A^{\mu}A_{\mu} -ie(\nabla_{\mu}A^{\mu})]\phi -2ieA^{\mu}D_{\mu}\phi.
\end{equation}
We see that $J$ transforms under the gauge transformations (\ref{q_g_transf}) 
in the same way as $\phi$, so (\ref{complex_J}) is covariant. Similarly, the
currents $j^{(\pm)}_{\mu}$ given by $(\ref{g28})$ are invariant with
respect to gauge transformations (\ref{q_g_transf}). The local particle 
production is described by the gauge-invariant generalization of
(\ref{complex_J_prod}). In a similar way as 
before, it can be shown that this local particle production is consistent with 
the particle production described in the usual way using the interaction 
picture, in which the interaction Hamiltonian is the part of the Hamiltonian 
that generates the right-hand side of (\ref{complex_J}).  

The conservation of charge is a little bit tricky. One can introduce 
two charge currents:
\begin{equation}
q_{\mu}=i\phi^{\dagger}\gapm \phi, \;\;\; 
\hat{q}_{\mu}=i\phi^{\dagger}\,\hat{\gapm}\, \phi.
\end{equation}
Again, it can be shown that
\begin{equation}
j^{(-)}_{\mu}=q_{\mu}-\langle 0|q_{\mu}|0\rangle=\, :\! q_{\mu}\! :.
\end{equation}
However, this current is not conserved. From the equation of motion 
(\ref{q_g_eq_m}) it follows that the current $\hat{q}_{\mu}$ is conserved, 
while $q_{\mu}$ is not:
\begin{eqnarray}\label{non-con-q}
& \nabla^{\mu}\hat{q}_{\mu} =0, & \nonumber \\
& \nabla^{\mu}{q}_{\mu} =2e[(\nabla_{\mu}A^{\mu})\phi^{\dagger}\phi+ 
A^{\mu}\nabla_{\mu}(\phi^{\dagger}\phi)]. &
\end{eqnarray}
Note that the right-hand sides in (\ref{non-con-q}) do not depend on 
the background $\bar{A}_{\mu}$, so the following peculiarities appear 
even when $\bar{A}_{\mu}=0$. The two currents are related as 
\begin{equation}\label{related_as}
\hat{q}_{\mu}={q}_{\mu}-2eA_{\mu}\phi^{\dagger}\phi.
\end{equation}
Which charge current is more physical, $\hat{q}_{\mu}$ or $q_{\mu}$? It 
depends on how the charge is measured.
If the charge is measured so that the 
corresponding electromagnetic effects determined by the total 
electromagnetic field $\hat{F}_{\mu\nu}=\partial_{\mu}\hat{A}_{\nu}-
\partial_{\nu}\hat{A}_{\mu}$ are measured, then the current  
$\hat{q}_{\mu}$ is the physical one, because the Maxwell equations read
\begin{equation}
\nabla^{\nu}\hat{F}_{\mu\nu}=-e:\!\hat{q}_{\mu}\! : .
\end{equation}
However, this does not mean that the current $q_{\mu}$ is unphysical. 
Each hermitian operator, conserved or not, is an observable. 
In order to understand the physical meaning of $q_{\mu}$, 
in the rest of this subsection we work in the interaction picture, in which 
the field $\phi$ has the expansion (\ref{gexp}) and $A_{\mu}$ has a 
similar expansion in terms of creation and annihilation operators. 
If one simply 
counts the number of particles and the number of antiparticles, then 
the current $q_{\mu}$ is the physical one because the corresponding charge   
\begin{equation}
Q=\cint^{\mu}:\! q_{\mu}\! :\, =N^{(-)}\equiv\sum_k a_k^{\dagger}a_k
-b_k^{\dagger}b_k
\end{equation}
is equal to the difference of these two numbers. On the other hand, the 
conserved charge 
\begin{equation}\label{hat_Q}
\hat{Q}=\cint^{\mu}:\!\hat{q}_{\mu}\! :\, =N^{(-)}-2e\cint^{\mu}A_{\mu}
:\!\phi^{\dagger}\phi\! : 
\end{equation}
depends not only on the number of charged 
particles, but also on the number of photons. 

An interesting question is the following. Can one observe, by counting 
the number of scalar particles and antiparticles, that the charge $Q$ 
is not conserved? Since the extra term in (\ref{related_as}) is linear 
in $A_{\mu}$, the expected value of this operator vanishes for 
any state in which the number of photons is definite (i.e., the state 
is not a superposition of states with different numbers of photons). 
Therefore, the expected values of the
charges $Q$ and $\hat{Q}$ are equal for such states, 
so the charge $Q$ is conserved in this case. Note also that the number 
of any kind of particles is definite in any particular Feynman diagram, so the 
charge $Q$ is also conserved on the level of Feynman diagrams. The 
conservation of $Q$ on the level of Feynman diagrams can also be seen
directly from the fact that, in the interaction Hamiltonian, the fields 
$\phi$ and $\phi^{\dagger}$ always appear in products of the form
$\phi^{\dagger}\phi$ and $\phi^{\dagger}\gapm\phi$. However, if the 
initial state $|i\rangle$ is a state with a definite number of all kinds 
of particles, then the final state determined by the $S$-matrix operator 
takes the form 
\begin{equation}
|f\rangle =S|i\rangle =\sum_{n} C_{if_n}|f_n\rangle. 
\end{equation}
Here the states $|f_n\rangle$ have a definite number of all kinds of 
particles, but the total state $|f\rangle$ has not. (The coefficients 
$C_{if_n}=\langle f_n|f\rangle=\langle f_n|S|i\rangle$ 
are the amplitudes determined by the corresponding Feynman 
diagrams.) Therefore, it is possible, in 
principle, to find the system in a state in which the final charge $Q$ 
is not equal to the initial charge $Q$. However, it seems very 
unlikely that this will happen in practice because $Q$ is 
conserved in any experiment in which the state $|f\rangle$ collapses to a 
state with a definite 
number of particles, or the decoherence destroys all mixing terms 
so that the observable degrees of freedom can be described by a density 
matrix that is not a pure state, but a {\em classical} mixture of 
the states with different numbers of particles.

Note, finally, that analogous problems with the conservation of charge do not 
exist for spin-1/2 fields. In the Standard Model, fundamental charged scalar 
fields do not exist. There is the complex Higgs field, but, 
after the spontaneous symmetry breaking, it reduces to an 
uncharged hermitian Higgs field. Therefore, the problems with the 
conservation of charge for complex scalar fields may be phenomenologically 
irrelevant.   

\section{Spinor field}\label{SEC5} 

For simplicity, we study the spinor fields in Minkowski spacetime.
However, all equations can be generalized 
to an arbitrary spacetime by replacing the derivatives $\partial_{\mu}$ 
with the general-covariant derivatives and by 
replacing certain constant Dirac matrices 
with the $x$-dependent Dirac matrices 
$\gamma_{\mu}(x)=e_{\mu}^{\alpha}(x)\gamma_{\alpha}$, where 
$e^{\alpha}_{\mu}(x)$ is the vierbein \cite{bd}. 
When appropriate, 
we shortly discuss the effects of this generalization.
 
\subsection{Particle and charge currents}
\label{SPINOR}

The Dirac matrices satisfy the anticommutation relations
\begin{equation}\label{dirac_alg}
\{\gamma_{\mu},\gamma_{\nu}\} =2\eta_{\mu\nu}.
\end{equation}
We choose their representation such that \cite{bd2}
\begin{equation}\label{gama_prop}
\gamma_{\mu}^{\dagger}=\gamma_0\gamma_{\mu}\gamma_0.
\end{equation}
The massive spin-1/2 field $\psi(x)$ satisfies the Dirac equation
\begin{equation}\label{D1}
(i\!\not\!\partial-m)\psi=0, \;\;\; 
\bar{\psi}(i\!\not\stackrel{\leftarrow}{\partial}+m)=0,
\end{equation}  
where $\not\!\partial=\gamma_{\mu}\partial^{\mu}$
and $\bar{\psi}=\psi^{\dagger}\gamma_0$.
The field can be expanded as
\begin{eqnarray}\label{D2}
& \psi(x)=\displaystyle\sum_k 
b_k u_k(x) + d_k^{\dagger} v_k(x), & \nonumber \\
& \bar{\psi}(x)=\displaystyle\sum_k 
b_k^{\dagger} \bar{u}_k(x) + d_k \bar{v}_k(x), &
\end{eqnarray}
where the summation 
over $k$ includes the summation over spin indices as well.
The scalar product is
\begin{equation}\label{D3} 
(\psi,\chi)=\cint^\mu \bar{\psi}\gamma_{\mu}\chi.
\end{equation}
The spinors $u_k$ and $v_k$ satisfy
\begin{eqnarray}\label{D4}
& (u_k,u_{k'})=(v_k,v_{k'})=\delta_{kk'}, & \nonumber \\
& (u_k,v_{k'})=(v_k,u_{k'})=0. &
\end{eqnarray}
Therefore,
\begin{eqnarray}\label{D5}
& b_k=(u_k,\psi), \;\;\; d_k^{\dagger}=(v_k,\psi), & \nonumber \\
& b_k^{\dagger}=(\psi,u_k), \;\;\; d_k=(\psi,v_k). &
\end{eqnarray}
The canonical anticommutation relations can be written as
\begin{eqnarray}\label{D6}
& \{ \psi_a(x),\psi_b(x') \}_{\Sigma}
= \{ (\bar{\psi}(x)\gamma_0)_a,(\bar{\psi}(x')\gamma_0)_b \}_{\Sigma} =0, 
& \nonumber \\
& \{ \psi_a(x),\bar{\psi}_c(x') \}
(\gamma_0)_{cb}|_{\Sigma}=\delta_{ab}\delta^3({\bf x}-{\bf x'}). &
\end{eqnarray}
From (\ref{D6}) it follows that the operators (\ref{D5}) satisfy 
the usual algebra of fermion creation and annihilation operators:
\begin{equation}\label{D7}
\{ b_k, b_{k'}^{\dagger} \} = \{ d_k, d_{k'}^{\dagger} \} =\delta_{kk'},
\end{equation}
while other anticommutators vanish.
We introduce the particle and antiparticle 2-point functions
\begin{eqnarray}\label{D8}
& S^{(P)}(x,x')=\displaystyle\sum_k u_k(x) \bar{u}_k(x'), & \nonumber \\
& S^{(A)}(x,x')=\displaystyle\sum_k v_k(x') \bar{v}_k(x). &
\end{eqnarray}
They satisfy
\begin{eqnarray}\label{D9}
& (i\!\not\!\partial-m)S^{(P)}(x,x')=0, \;\;\;
S^{(P)}(x,x')(i\!\not\,\stackrel{\leftarrow}{\partial'}+m)=0, & \nonumber \\
& S^{(A)}(x,x')(i\!\not\stackrel{\leftarrow}{\partial}+m)=0, \;\;\;
(i\!\not\!\partial'-m)S^{(A)}(x,x')=0 &
\end{eqnarray}
and have the property
\begin{eqnarray}\label{D10}
& (S^{(P)}(x,x'))_{ab}=\langle 0|\psi_a(x)\bar{\psi}_b(x')|0\rangle, 
& \nonumber \\
& (S^{(A)}(x,x'))_{ab}=\langle 0|\bar{\psi}_b(x)\psi_a(x')|0\rangle. &
\end{eqnarray}
We also introduce the 2-point functions $S^{(P)}_{\mu\nu}(x,x')$ and 
$S^{(A)}_{\mu\nu}(x,x')$ defined as
\begin{equation}\label{D11}
S^{(P,A)}_{\mu\nu}(x,x')=\gamma_{\mu}S^{(P,A)}(x,x')\gamma_{\nu}.
\end{equation}
Using (\ref{dirac_alg}), (\ref{gama_prop}) and (\ref{D8}) we find
\begin{equation}\label{D12}
\gamma_0 S^{(P,A)\dagger}(x,x')\gamma_0=S^{(P,A)}(x',x),
\end{equation}
while from (\ref{dirac_alg}), (\ref{gama_prop}), (\ref{D6}), and  
(\ref{D10}) it follows that
\begin{eqnarray}\label{D13}
& [S^{(P)}(x,x')-S^{(A)}(x',x)]_{\Sigma}=\gamma_0 
\delta^3({\bf x}-{\bf x'}), & \nonumber \\
& [S^{(P)}_{\mu\nu}(x,x')-S^{(A)}_{\mu\nu}(x',x)]_{\Sigma}=
\gamma_{\mu}\gamma_0\gamma_{\nu} 
\delta^3({\bf x}-{\bf x'}), & \nonumber \\
& [S^{(P)}_{\mu 0}(x,x')-S^{(A)}_{\mu 0}(x',x)]_{\Sigma}=
\gamma_{\mu}\delta^3({\bf x}-{\bf x'}). &   
\end{eqnarray}

The global quantities
\begin{equation}\label{D14}
N^{(\pm)}=\sum_k b_k^{\dagger}b_k \pm d_k^{\dagger}d_k
\end{equation}
can be written in the covariant form as
\begin{equation}
N^{(\pm)}=\cint^{\mu}j^{(\pm)}_{\mu},
\end{equation}
where
\begin{eqnarray}\label{D15}
j^{(\pm)}_{\mu}(x) & = & \cint'^{\nu} \frac{1}{2} \{ 
[\bar{\psi}(x)S^{(P)}_{\mu\nu}(x,x')\psi(x') \nonumber \\ 
& & \pm  
\psi^T(x)S^{(A)T}_{\nu\mu}(x,x')\bar{\psi}^T(x')] +{\rm h.c.} \}
\end{eqnarray} 
and the superscript $T$ denotes the transpose.
From (\ref{D1}) and (\ref{D9}) it follows that both currents are 
conserved:
\begin{equation}
\partial^{\mu}j^{(\pm)}_{\mu}=0.
\end{equation}

By an appropriate generalization to a curved background, 
it is straightforward to see that the currents $j^{(\pm)}_{\mu}$ are 
conserved again. Also, one can introduce different natural spinors 
at different times, which redefines the 2-point functions (\ref{D8}) 
and leads to a local description of particle production by the gravitational 
background consistent with the Bogoliubov transformation method.
The most natural choice of the 2-point functions is
\begin{eqnarray}\label{D16}
& S^{(P)}(x,x')=(i\!\not\!\nabla+m)\bar{G}^+(x,x'), & \nonumber \\
& S^{(A)}(x,x')=\bar{G}^+(x,x')(i\!\not\stackrel{\leftarrow}{\nabla}-m), &
\end{eqnarray}
where $\bar{G}^+(x,x')$ is the 2-point function for scalar fields 
with $\xi=1/4$ \cite{bd} 
discussed in Sec.~\ref{NATURAL}. This natural choice also leads 
to the conservation of the particle current. 

Using (\ref{D6}) and (\ref{D13}) we find the standard form of the 
charge current
\begin{equation}\label{D17}
j^{(-)}_{\mu}(x)=\bar{\psi}(x)\gamma_{\mu}\psi(x)-
{\rm Tr}S^{(A)}(x,x)\gamma_{\mu},
\end{equation}
which, after using (\ref{D2}) and (\ref{D8}), can be written as 
\begin{equation}\label{D17.1}
j^{(-)}_{\mu}(x)=\bar{\psi}(x)\gamma_{\mu}\psi(x)-
\langle 0|\bar{\psi}(x)\gamma_{\mu}\psi(x) |0\rangle.
\end{equation}

The currents $j^{(\pm)}_{\mu}$ can also be written as
\begin{equation}
j^{(\pm)}_{\mu}=j^{(P)}_{\mu} \pm j^{(A)}_{\mu}.
\end{equation}
In terms of the creation and annihilation operators, the currents can be 
written as
\begin{equation}
j^{(P)}_{\mu}=\sum_{k,k'}
\bar{u}_{k'}\gamma_{\mu}u_k b_{k'}^{\dagger}b_k +j^{{\rm mix}}_{\mu},
\end{equation}
\begin{equation}
j^{(A)}_{\mu}=\sum_{k,k'}
\bar{v}_k\gamma_{\mu}v_{k'} d_{k'}^{\dagger}d_k -j^{{\rm mix}}_{\mu},
\end{equation}
\begin{equation}
j^{(+)}_{\mu}=\sum_{k,k'}
\bar{u}_{k'}\gamma_{\mu}u_k b_{k'}^{\dagger}b_k +
\bar{v}_k\gamma_{\mu}v_{k'} d_{k'}^{\dagger}d_k,
\end{equation}
\begin{equation}                                 
j^{(-)}_{\mu}=\sum_{k,k'}(
\bar{u}_{k'}\gamma_{\mu}u_k b_{k'}^{\dagger}b_k -
\bar{v}_k\gamma_{\mu}v_{k'} d_{k'}^{\dagger}d_k) +2j^{{\rm mix}}_{\mu},
\end{equation}
where
\begin{equation}
j^{{\rm mix}}_{\mu}=\frac{1}{2}\sum_{k,k'} (
\bar{v}_k\gamma_{\mu}u_{k'} d_k b_{k'} + 
\bar{u}_{k'}\gamma_{\mu}v_k b_{k'}^{\dagger}d_k^{\dagger}).
\end{equation} 

\subsection{The generalization to the electromagnetic interaction}

The expressions for the currents in Sec.~\ref{SPINOR} are already 
gauge invariant. This makes the generalization to the case of 
electromagnetic interaction easier than for the scalar fields. 
This is related to the fact that the Dirac equation is a 
first-order equation, so the expressions for the currents contain 
$\gamma_{\mu}$ instead of $\partial_{\mu}$. 
Consequently, in the currents, one does not 
need to replace the derivatives with the gauge-covariant derivatives. 
Therefore, the currents $j^{(\pm)}_{\mu}$ are given by the 
expression (\ref{D15}). However, the fields satisfy
\begin{equation}\label{gD1}
(i\!\not\!\hat{D} -m)\psi=0, \;\;\;
\bar{\psi}(i\!\not\!\hat{\stackrel{\leftarrow\;}{D^*}}+m)=0,
\end{equation}
while the 2-point functions satisfy 
\begin{eqnarray}\label{gD9}
& (i\!\not\!D -m)S^{(P)}(x,x')=0, \;\;\;
S^{(P)}(x,x')(i\!\not\stackrel{\leftarrow\;\;}{D'^*}+m)=0, & \nonumber \\
& S^{(A)}(x,x')(i\!\not\stackrel{\leftarrow\;}{D^*}+m)=0, \;\;\;
(i\!\not\!D'-m)S^{(A)}(x,x')=0. &
\end{eqnarray}
The charge current can be written in the form (\ref{D17.1}) again 
and is conserved. On the other hand, the particle current is not 
conserved:
\begin{eqnarray}\label{ppgs}
\partial^{\mu}j^{(+)}_{\mu}(x) & = &
\cint'^{\nu} \frac{1}{2} \{ ieA_{\mu}(x)
[\bar{\psi}(x)S^{(P)}_{\mu\nu}(x,x')\psi(x') \nonumber \\
& & -
\psi^T(x)S^{(A)T}_{\nu\mu}(x,x')\bar{\psi}^T(x')] +{\rm h.c.} \}.
\end{eqnarray}
The background $\bar{A}_{\mu}$ does not appear in (\ref{ppgs}), 
which means that the background electromagnetic field does not 
produce particles. Again, it may be interpreted as a consequence 
of the existence of the preferred spinors $v_k$ and $u_k$, or as a 
consequence of the existence of the natural 2-point functions 
calculated by using Schwinger's method \cite{schw,tsam}. The 
particle production by the electromagnetic background can be described 
by introducing different natural spinors
at different times, which redefines the 2-point functions
and leads to a local description of particle production 
consistent with the Bogoliubov transformation method. The particle 
production described by (\ref{ppgs}) is equivalent to the usual 
description based on the interaction picture.

\section{Conclusion and outlook}\label{SEC6}  

In this paper, the currents of particles and charge 
for scalar and spinor fields in gravitational and electromagnetic 
backgrounds have been constructed. 
The currents are covariant with respect to general coordinate 
transformations and invariant with respect to gauge transformations. 
The currents of charge constructed using our method are equal to the 
usual charge currents. However, previous definitions of particles 
in quantum field theory were not general-covariant and 
gauge-invariant. This is because previous definitions of particles 
were not local, i.e., the operator of the local particle density 
was not known, except for the non-relativistic case. 

For a given background, the particle current is not unique, but
depends on the choice of a 2-point function.
Different choices correspond to different definitions of particles.
There are 3 types of this choice. The first type is based on the 
choice of a particular set of complete orthonormal modes 
that satisfy the field equations of motion. The resulting 
particle currents give a local description of the particle content in various
inequivalent representations of field algebra. The second type 
is based on choosing different natural modes at different times. 
This leads to a local description of particle production by classical 
backgrounds. When the total number of particles on a 
Cauchy hypersurface is calculated, then the first and second types 
of choice lead to two types of conventional 
global concepts of particles. 
However, the third type, based on the 2-point function calculated using 
Schwinger's method, is novel even on the global level. This type 
seems to be the most natural one. It is tempting to interpret these
particles as real particles.
If this interpretation is correct, then
classical backgrounds do not produce real particles.
There are also other indications that classical
backgrounds might not produce particles 
\cite{padmprl,belin,nikol1,nikol2,nikol3}.
Even if these particles 
do not correspond to real physical particles in general,
it is interesting to ask about the physical meaning of this
hermitian observable that corresponds to physical
particles at least in Minkowski spacetime and zero electromagnetic 
background.

As by-products, relevant even without our novel description 
of particles, the following results
have also been obtained. A unique 
Green function $\bar{G}(x,x')$ has been constructed for the case 
in which $x$ and $x'$ can be connected by more than one geodesic. 
It has been found that the requirement of covariance of the time 
evolution leads to a preferred role of Gaussian coordinates. 
The peculiarities related to the concept of charge and its conservation 
in scalar QED have been clarified.    

In this paper, the phenomenological applications 
of the developed formalism have not been studied, 
but we hope that our results will 
motivate further investigations. We suggest several possible applications.

First, one can adopt the variant of the formalism according to which 
background gravitational and electromagnetic fields do 
produce particles. Contrary to the previous methods, our method 
enables one to calculate {\em where} particles are created. This question 
is particularly interesting in the case of particle production by 
black holes. Our formalism is able to test the usual conjecture 
that particles are created near the horizon.

Second, one can adopt a variant of the formalism according to which 
background gravitational and electromagnetic fields do 
not produce particles. It seems very likely 
that this is the correct interpretation because it results 
from the definition of particles that seems to be the most natural.
In order to calculate particle currents or the total number of 
particles on a spacelike Cauchy hypersurface, 
one has to calculate 2-point functions explicitly by using 
the Schwinger-DeWitt method, which is a 
non-trivial task in general.

Third, one can study possible physical implications of the fact that 
the local particle density may be negative. 
It would be interesting 
to study more thoroughly how the appearance of negative particle 
densities depends on the state, entanglement and 
background interactions, as well as
how negative densities
disappear by the wave-function collapse or decoherence. 
For example, the appearance of negative particle 
densities could be related to the existence of classical ``barriers" such as 
a horizon or an electric potential higher than $2m$. Eq. 
(\ref{22_mu}) suggests that negative particle densities may be 
related to positive-norm solutions that contain both the positive and the
negative frequencies, but this depends 
on which of the 3 types of 2-point functions is adopted.
 
In any case, we believe that our results could influence further research
that could result in a significant change of the understanding 
of the concept of particles in quantum field theory. Our local approach 
changes the notion of a relativistic particle even for a free particle 
in Minkowski spacetime and zero electromagnetic background.

\section*{Acknowledgments} 
The author is grateful to S.~A.~Fulling for a discussion, especially 
on the possible preferred role of Gaussian coordinates in 
curved-spacetime quantum field theory. 
This work was supported by the Ministry of Science and Technology of the
Republic of Croatia under Contract No.~0098002.

\appendix
\section{Particle density in non-relativistic quantum field theory}

In non-relativistic quantum field theory, the operator of particle 
density is a well-known quantity \cite{schiff,schweber}. This is a 
non-negative operator
\begin{equation}\label{A13}
n(x)=\Psi^{\dagger}(x)\Psi(x),
\end{equation}
where $x=({\bf x},t)$ and $\Psi$ satisfies the Schr\"{o}dinger equation
\begin{equation}\label{A1}
-\frac{1}{2m}\nabla^2\Psi +U\Psi =i\frac{\partial\Psi}{\partial t}, 
\end{equation}
with $U(x)$ being a non-relativistic potential. 
In this appendix, we rederive (\ref{A13}) using the method employed in 
this paper.  

Eq. (\ref{A1}) can be derived from the Lagrangian density
\begin{equation}
{\cal L}=i\Psi^{\dagger}\frac{\partial\Psi}{\partial t} -\frac{1}{2m}
(\nabla\Psi^{\dagger})(\nabla\Psi) -U\Psi^{\dagger}\Psi.
\end{equation}
Therefore, the canonical momentum is $\pi=i\Psi^{\dagger}$ and the 
canonical equal-time commutation relation between $\Psi$ and $\pi$ 
can be written as
\begin{equation}\label{Acomm}
[\Psi({\bf x},t),\Psi^{\dagger}({\bf x}',t)]=\delta^3({\bf x}-{\bf x}').
\end{equation}
The scalar product $(\phi_1,\phi_2)=\int d^3x \,\phi^*_1\phi_2$ 
does not depend 
on time, provided that $\phi_1$ and $\phi_2$ satisfy (\ref{A1}). 
The fields $\Psi$ and $\Psi^{\dagger}$ have the expansion
\begin{equation}\label{A4}
\Psi(x)=\sum_k a_k u_k(x), \;\;\; 
\Psi^{\dagger}(x)=\sum_k a^{\dagger}_k u^*_k(x).
\end{equation}
The modes $u_k$ are orthogonal:
\begin{equation}\label{A5}
(u_k,u_{k'})=(u^*_k,u^*_{k'})=\delta_{kk'}.
\end{equation}
From (\ref{A4}) and (\ref{A5}) we find
\begin{equation}\label{A9}
a_k=(u_k,\Psi), \;\;\; a^{\dagger}_k=(u^*_k,\Psi^{\dagger}).
\end{equation}
From the equal-time commutation relations it follows that $a_k$ and
$a^{\dagger}_k$ satisfy the usual algebra of lowering and raising 
operators \cite{schiff}, so the operator
\begin{equation}\label{A8}
N=\sum_k a^{\dagger}_k a_k
\end{equation}
is the operator of the number of particles. 
From (\ref{A4}) we see that 
\begin{equation}\label{A12pom}
\langle 0| \Psi^{\dagger}(x)\Psi(x') |0\rangle =0, \;\;\; 
\langle 0| \Psi(x)\Psi^{\dagger}(x') |0\rangle =W(x,x'),
\end{equation}
where
\begin{equation}\label{A12}
W(x,x')=\sum_k u_k(x) u^*_k(x').
\end{equation}
From (\ref{Acomm}), (\ref{A12pom}), and (\ref{A12}) we see that the 
modes $u_k$ are complete, i.e., that
\begin{equation}\label{A6}
W(({\bf x},t), ({\bf x}',t))=\delta^3({\bf x}-{\bf x}').
\end{equation}

Using (\ref{A9}) and (\ref{A12}), we can 
write (\ref{A8}) as
\begin{equation}\label{A10}
N=\int d^3x \, n(x),
\end{equation}
where
\begin{equation}\label{A11}
n(x)=\int d^3x' \frac{1}{2} \{ W(x,x')\Psi^{\dagger}(x)\Psi(x')
+ {\rm h.c.} \}.
\end{equation}
In (\ref{A11}), $t=t'$, so (\ref{A11}) and (\ref{A6}) lead to 
(\ref{A13}). Therefore, contrary to the relativistic case, 
in the non-relativistic case the 
particle density is non-negative and can be written in a purely local 
form. 

For comparison with the relativistic case, let us study the free case 
$U=0$. The modes $u_k$ can be chosen to be the plane-wave modes 
$u_{{\bf k}}$, which, in a finite volume $V$, are
\begin{equation}
u_{{\bf k}}({\bf x},t)=\frac{e^{-i(E_{{\bf k}}t-{\bf k}{\bf x})}}
{\sqrt{V}},
\end{equation}
where $E_{{\bf k}}={\bf k}^2/2m$. In particular, taking the state to be  
$|\psi\rangle =2^{-1/2}(|{\bf q}_1\rangle + |{\bf q}_2\rangle)$, 
which is the non-relativistic analog of the state  
$|\psi\rangle$ in Eq. (\ref{number53}), 
we find
\begin{equation}\label{number53nr}
\langle\psi|n|\psi\rangle =V^{-1} \{ 1+
{\rm cos}[ (E_{{\bf q}_1}-E_{{\bf q}_2})t - 
({\bf q}_1-{\bf q}_2){\bf x} ] \} .
\end{equation}
Contrary to the relativistic particle density (\ref{number53}), 
the non-relativistic particle density (\ref{number53nr}) is 
non-negative.

\end{document}